# In situ Investigation of Neutrals Involved in the Formation of Titan Tholins

David Dubois[1], Nathalie Carrasco[1,2], Marie Petrucciani[1], Ludovic Vettier[1], Sarah Tigrine[1,3], Pascal Pernot[4]

[1] LATMOS Laboratoire Atmosphères, Milieux et Observations Spatiales, Université Versailles St-Quentin, 78280 Guyancourt, France

[2] Institut Universitaire de France, Paris F-75005, France

[3] Synchrotron SOLEIL, l'Orme des Merisiers, St Aubin, BP 48, 91192 Gif-sur-Yvette Cedex, France

[4] Laboratoire de Chimie Physique, UMR 8000, CNRS, Université Paris-Sud 11, 91405, Orsay, France

**Abstract**

The Cassini Mission has greatly improved our understanding of the dynamics and chemical processes occurring in Titan's atmosphere. It has also provided us with more insight into the formation of the aerosols in the upper atmospheric layers.

However, the chemical composition and mechanisms leading to their formation were out of reach for the instruments onboard Cassini. In this context, it is deemed necessary to apply and exploit laboratory simulations to better understand the chemical reactivity occurring in the gas phase of Titan-like conditions. In the present work, we report gas phase results obtained from a plasma discharge simulating the chemical processes in Titan's ionosphere. We use the PAMPRE cold dusty plasma experiment with an $N_2$-$CH_4$ gaseous mixture under controlled pressure and gas influx. An internal cryogenic trap has been developed to accumulate the gas products during their production and facilitate their detection. The cryogenic trap condenses the gas-phase precursors while they are forming, so that aerosols are no longer observed during the 2h plasma discharge. We focus mainly on neutral products $NH_3$, HCN, $C_2H_2$ and $C_2H_4$. The latter are identified and quantified by *in situ* mass spectrometry and infrared spectroscopy. We present here results from this experiment with mixing ratios of 90-10% and 99-1% $N_2$-$CH_4$, covering the range of methane concentrations encountered in Titan's ionosphere. We also detect *in situ* heavy molecules ($C_7$). In particular, we show the role of ethylene and other volatiles as key solid-phase precursors.

**Keywords:** Titan, Saturn; Atmospheres, chemistry; Plasma discharge; Experimental techniques

# I) Introduction

Titan is Saturn's largest moon and second largest in the Solar System after Ganymede. It has a mean radius of 2,575 km. One major interest in Titan is its uniquely dense (a 1.5-bar ground pressure) and extensive (~1,500 km) atmosphere, primarily made of $N_2$ and $CH_4$. In fact, the extent of the atmosphere alone accounts for about 58% in size of Titan's radius. Titan's first flyby by Pioneer 11 (1979) revealed the moon as being a faint brownish dot orbiting Saturn. The subsequent Voyager 1 (1980), Voyager 2 (1981), Cassini (2004 - 2017) and Huygens (2005) spacecraft have uncovered physicochemical processes unique in the solar system occurring both in the atmosphere and surface. Early seminal studies (Khare and Sagan, 1973; Sagan, 1973; Hanel et al., 1981) showed that Titan's reducing atmosphere is also composed of complex and heavy organic molecules and aerosols, their laboratory analogous counterparts coined *tholins*. This chemistry is initiated at high altitudes and participates in the formation of solid organic particles (Khare et al., 1981; Waite et al., 2007, and Hörst, 2017 for a more recent and holistic review of Titan's atmospheric chemistry). At altitudes higher than 800 km, the atmosphere is under the influence of energy deposition such as solar harsh ultraviolet (VUV) radiation, solar X-rays, Saturn's magnetospheric energetic electrons and solar wind (Krasnopolsky, 2009; Sittler et al., 2009). Cosmic dust input is also thought to chemically interact with volatiles in the middle atmosphere (English et al., 1996; Frankland et al., 2016). To wit, Titan is presumably one of the most chemically complex bodies in the solar system. Some of the techniques used to better understand the processes occurring in Titan's atmosphere are detailed hereafter.

Earth-based observations of Titan have been one way to determine the atmospheric composition and structure, as well as surface and subsurface features. For example, direct monitoring of seasonal tropospheric clouds from Earth-based observations helped us infer their transient locations (Brown et al., 2002). Ground-based submillimeter observations were used to infer the stratospheric composition with for example, the determination of hydrogen cyanide (HCN) and acetonitrile ($CH_3CN$) mixing ratios (Tanguy et al., 1990; Bezard and Paubert, 1993; Hidayat et al., 1997), and with the rise of ground-based observation techniques such as the Atacama Large Millimeter Array (ALMA) (Cordiner et al., 2014a, 2014b; Molter et al., 2016; Serigano et al., 2016; Palmer et al., 2017). The use of other space-based telescopes such as the Infrared Space Observatory (ISO) and the Herschel Space Observatory has additionally provided information on the stratospheric composition and neutral abundances (Coustenis et al., 2003; Courtin et al., 2011; Rengel et al., 2014). Early on, atmospheric studies brought by the Voyager 1 mission were able to identify several organic molecules (Hanel et al., 1981; Kunde et al., 1981; Maguire et al., 1981). Later, the Cassini-Huygens (NASA/ESA) mission provided us with a more precise atmospheric characterization of the composition (Teanby et al., 2012; Vinatier et al., 2015; Coustenis et al., 2016), wind (Flasar et al., 2005; Achterberg et al., 2011) and upper atmosphere thermal profiles (e.g. Snowden et al., 2013), while the Huygens lander (ESA) was able to provide insight on the *in situ* composition (Israel et al., 2005) and optical properties of the aerosols (Doose et al., 2016). Cassini's Composite Infrared Spectrometer (CIRS) limb spectral imaging (Vinatier et al., 2009) examined the role of volatiles in Titan's chemistry and atmospheric dynamics. Many hydrocarbons and nitriles were shown to have wide latitudinal variations of vertical mixing ratios (sometimes contradicting Global Circulation Models, e.g. de Kok et al., 2014), emphasizing the importance of understanding the complex coupling between gas phase chemistry and global dynamics. Both polar enrichments of nitriles (e.g. Teanby et al., 2009) as well as the presence of heavy ions (Crary et al., 2009; Wellbrock et al., 2013; Desai et al., 2017) have also been suggested to participate in Titan's aerosol production. Thus, ground- and space-based observations have and will be especially important in the post-Cassini era (Nixon et al., 2016).

Photochemical models have been complementarily used to predict the thermal, compositional and structural evolution of Titan's atmospheric column (e.g. Atreya et al., 1978; Carrasco et al., 2007, 2008). Furthermore, these models offer us testable hypotheses for experiments and predictions for chemical pathways leading to the formation of aerosols (e.g. Waite et al., 2007; Yelle et al., 2010).

Lastly, experimental techniques have also been used to simulate Titan's atmospheric conditions (Khare et al., 1984; Sagan and Thompson, 1984; Pintassilgo et al., 1999; Cable et al., 2012; Carrasco et al., 2013; Sciamma-O'Brien et al., 2015, Tigrine et al., 2016; Hörst et al., 2018) and determine chemical mechanisms that are otherwise out of Cassini's reach. A wide range of laboratory simulations have been used to complementarily investigate Titan's atmospheric chemistry and different energy inputs (see Cable et al., 2012 for a thorough review of these experiments). Typically in our lab, plasma discharges (Szopa et al., 2006; Sciamma-O'Brien et al., 2010; Carrasco et al., 2012)  or UV irradiation (Peng et al., 2013) have been used to achieve Titan-like ionospheric conditions. In this context, it is deemed necessary to apply and exploit such a technique to better understand the chemical reactivity occurring in Titan-like upper atmospheric conditions. Hence, the gas phase which we know is key to the formation of aerosols on Titan (Waite et al., 2007) is emphasized. Indeed, chemical precursors that are present in the gas phase (Carrasco et al., 2012) follow certain chemical pathways that still need to be investigated. For instance, Sciamma-O'Brien et al., 2010 showed how the production of atomic hydrogen coming from $CH_4$ was anti-correlated with aerosol formation, by changing the initial methane amount in the discharge. They argued that the presence of molecular and atomic hydrogen in too large amounts in the gas mixture may have an inhibiting effect on the polymerization reaction that produce the tholins.

To simulate the methane mixing ratios and energy conditions present in Titan's ionosphere, we used the PAMPRE cold dusty plasma experiment (Szopa et al., 2006) with an $N_2$-$CH_4$ gas mixture under controlled gas influx and pressure. Cold plasma discharges have been shown to faithfully simulate the electron energy distribution of magnetospheric electrons and protons, and UV irradiation deposited at Titan's atmosphere within a range of ~ 10 eV – 5 keV (Sittler et al., 1983; Chang et al., 1991; Maurice et al., 1996; Brown et al., 2009; Johnson et al., 2016), making them good laboratory analogs to simulate the ionospheric chemistry.

A first study (Gautier et al., 2011) dealt with the identification of gas products in the PAMPRE reactor and used an external cold trap to study the gas phase. Analyses were made by gas-chromatography coupled to mass spectrometry. More than 30 reaction products were detected and a semi-quantitative study was performed on nitriles, the most abundant molecules collected. However, one of the limitations to this technique is how the products are transported *ex situ*, and may allow for loss or bias in their chemical reactivity and measurements. *In situ* detections were then made by mass spectrometry with no cryogenic trapping (Carrasco et al., 2012). Unfortunately, the only use of mass spectrometry prevented any univocal identification of large organic molecules likely involved as precursors for aerosol formation. Since then, setup improvements such as an internal cryogenic trap and infrared spectroscopy have been made in order to study the gas phase *in situ* using these different complementary techniques. The purpose of the cryotrap is to act as a cold finger enabling the volatiles to condensate and preventing them from being evacuated from the chamber. This has allowed us to avoid any volatile loss and to take on a quantitative approach on volatile products. Thus, to better characterize the initial volatile products precursors to aerosols, we used a coupled analysis of mass and infrared spectrometry. The presence of an internal cold trap made the condensation of these volatiles possible, allowing for *in situ* measurements.

We will first describe our experimental chamber, which was designed to be relevant to simulate Titan's ionosphere conditions. This was achieved by choosing methane concentrations pertaining to those found in the upper atmosphere. Mass spectrometry results of volatiles at two methane concentrations will be then presented, as well as their infrared spectroscopic analyses with a quantitative approach taken on four major neutral volatiles ($NH_3$, $C_2H_2$, $C_2H_4$ and HCN) acting as precursors to the formation of aerosols. Finally, we will discuss the discrepancies between both methane conditions, as well as ethylene, ammonia and hydrogen cyanide chemistry.

## II) Experimental Setup

### 2.1. The PAMPRE experiment and in situ cryogenic analysis

The PAMPRE cold plasma reactor used in this study aims, by design, at reproducing exogenic energy sources, for example Saturn's magnetosphere particles, UV radiation, cosmic rays and coronal processes impacting Titan's atmosphere (Szopa et al., 2006; Alcouffe et al., 2010). The reactor delivers a capacitively coupled plasma discharge, produced by a 13.56 MHz radiofrequency generator through a tuning match box. The gas influx is held constant and monitored with mass flow controllers. The latter can tune an $N_2$-$CH_4$ mixing ratio from 0% to 10% of $CH_4$ from highly pure $N_2$ (>99.999%) and 90-10% $N_2$-$CH_4$ (>99.999%) bottles. An *in situ* mass spectrometer (MS) and Fourier-Transform Infrared (FT-IR) spectrometer are also integrated to the system.
To simulate Titan's atmospheric conditions, 90-10% and 99-1% $N_2$-$CH_4$ ratios were considered. Both correspond to a 55 sccm flow in standard conditions. Pressure gauges are set at 0.9 mbar of constant pressure in the reactor while the plasma power is constant at 30W.

In order to detect, analyze and preserve all of the products during our experiments, we need a trapping mechanism to capture these compounds, and thus prevent them from being evacuated. Gautier et al. (2011) used an external cryogenic trap system, shaped as a cylindrical coil, immersed in liquid nitrogen. Then, they isolated the coiled trap and analyzed the gas phase *ex situ*. Here, the setup is different, and we believe more efficient as the trapping and subsequent analyses are performed *in situ* and avoid any transfer line bias. Moreover, this cryo-trap mainly traps the gas-phase chemistry in the preceding stages to the formation of solid organic aerosols, preventing their production. In the absence of such a cryo-trap, the plasma is dusty and several mg/h of solid particles are produced. As a result, this enables us to effectively study in detail these gas-phase precursors of tholins, trapped and accumulated during a 2h plasma discharge. Fig. 1 shows the 3D schematics of our cryogenic trap. An external liquid nitrogen tank extends as a copper circuit into the chamber for cooling. This cooling circuit goes through the upper lid of the chamber, and spirals down onto and lays flat on the polarized electrode. This cooling system is semi-closed, in that the remaining liquid nitrogen is stored in another container outside the reactor. The electrode temperature is controlled with a temperature regulator (Cryo-Diffusion Pt100 probe) which controls the liquid nitrogen circulation in the cooling circuit with a solenoid valve. The initial temperature is adjusted to -180 ±2°C prior to the experiment.

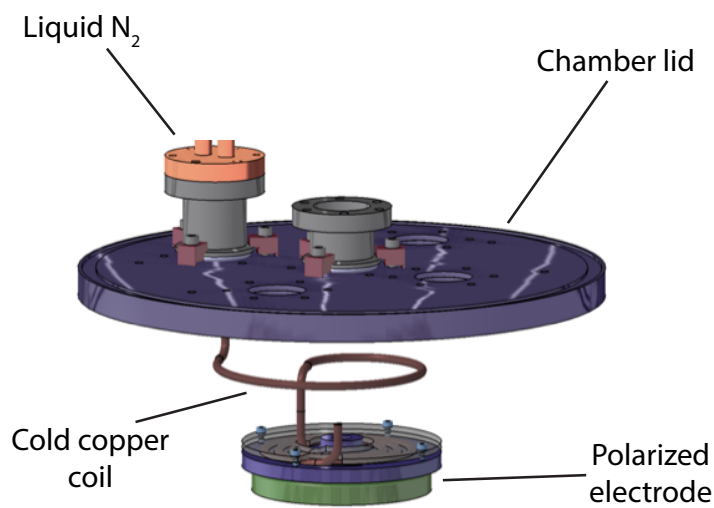

**Fig. 1.** 3D schematics of the cryogenic trap. The cold trap goes through the top lid of the reactor and coils down into the chamber, which is then in contact with the polarized electrode. The electrode is thus cooled to low-controlled temperatures and the liquid nitrogen goes out through an exit system. This cryogenic trap enables the *in situ* condensation of the produced volatiles within the plasma.

Using the cryogenic trap set at T = -180 ±2 ºC, the protocol used was the following: The plasma discharge and cooling system are turned on and the electrodes are kept for 2h at T = -180ºC. The duration of the plasma discharge was chosen to trap and accumulate the gas phase products, and improve their detection. The discharge reaches a steady state after a few minutes, as shown in Alcouffe et al., (2010). A similar 2h trapping was previously performed in Gautier et al., (2011) with an external cold trap, which was then analyzed by GC-MS. For consistency, we have chosen the same discharge duration. During this time, the products formed in the $N_2$-$CH_4$ plasma are accumulated and adsorbed onto the chamber walls still at low-controlled temperatures thanks to the cryogenic system. After 2h, the plasma is switched off, the chamber is pumped down from 0.9 mbar to $10^{-4}$ mbar, and the cold trap is stopped. The volatile products are gradually released under vacuum. We then proceed to do *in situ* gas phase measurements as the chamber goes back to room temperature (Fig. 3). During this temperature increase, it is possible that some additional neutral-neutral chemistry could occur. Our analytical capacity enables us to address the most abundant and therefore stable molecules, which are not expected to vary strongly apart from the sublimation process.

### *2.2. Mass Spectrometry*

QMS quadrupole mass spectrometers were used (Pfeiffer and Hiden Analytics) for *in situ* measurements inside the chamber. Once the neutral gas is ionized within the MS chamber (SEM voltage of 1000V), each mass is collected by the detector. The MS samples near the reactive plasma with a 0.8 mm-diameter capillary tube, long enough to keep a high-pressure gradient between the MS and plasma chambers. Thus, an effective $10^{-5}$-$10^{-6}$ mbar resident pressure in the MS chamber is maintained throughout the experiment. This vacuum background was thoroughly checked before each simulation. The experiments were done using a 1u resolution, covering a 1-100 mass range, at a speed of 2s/mass with a channeltron-type detection. Finally, a baseline noise filter was applied to each spectrum.

### 2.3. Infrared Spectroscopy

A ThermoFisher Nicolet 6700 Fourier Transform Infrared (FT-IR) spectrometer was used for *in situ* analysis of the gas phase in transmission mode. The spectrometer uses a Mercury Cadmium Telluride (MCT/A) high sensitivity detector with a KBr beamsplitter. During each measurement, 500 scans were taken at a resolution of 1 cm$^{-1}$ for 650–4000 cm$^{-1}$ corresponding to a short-wavelength mid-IR (15.4–2.5 μm) range. The optical path length is 30 cm.

## III) Results

We will now present results from two sets of experiments done at two initial methane concentrations, 1% and 10% (three and two replicates, respectively), representative of the methane concentration range found in Titan's ionosphere (Waite et al., 2005). The effect of the cold trap on the plasma can be appreciated by comparing the mass spectra in the same conditions with and without the cold trap (Fig. 2).

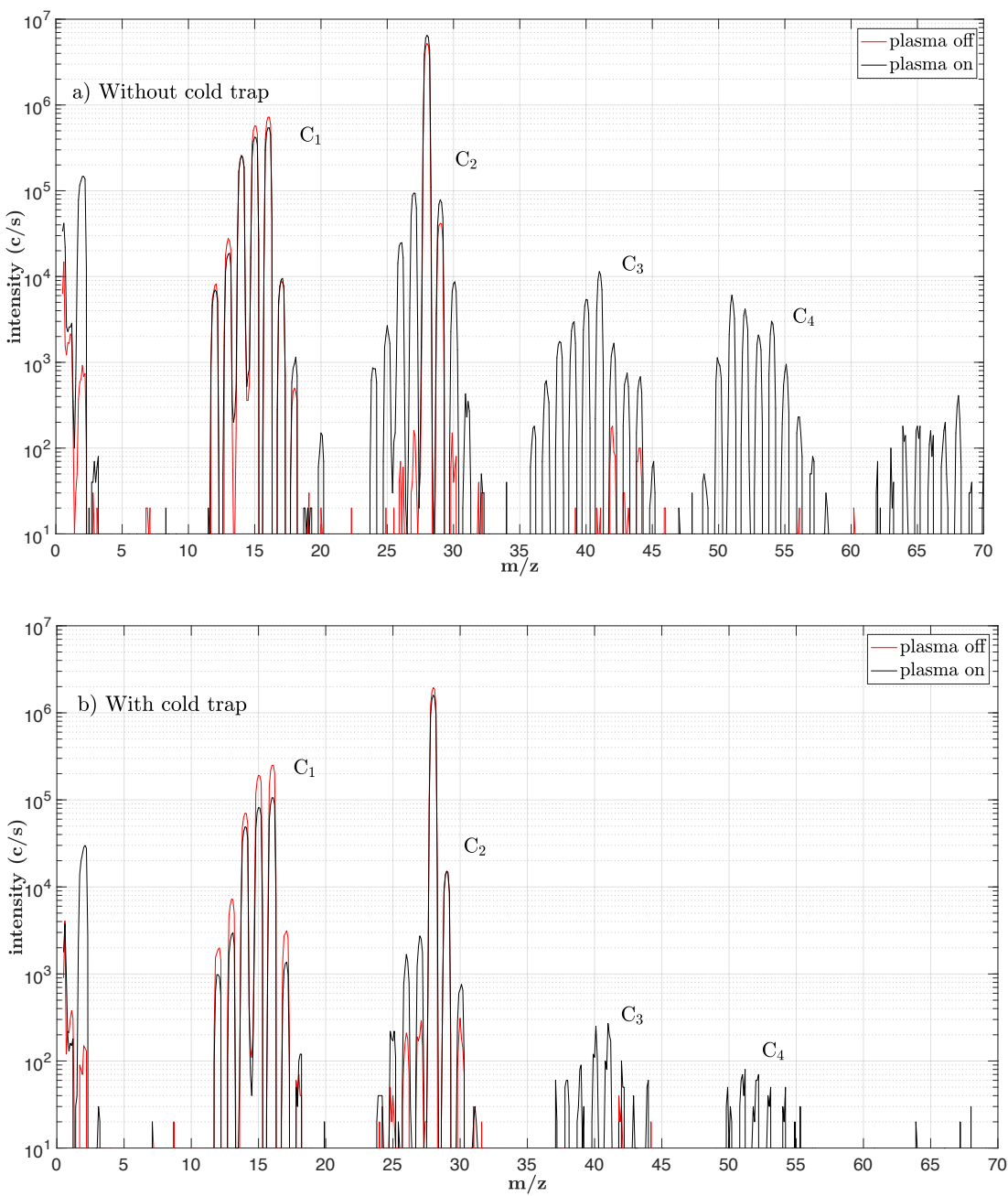

**Fig. 2.** (a) Upper panel: No cryogenic trap. Mass spectra of a 90-10% $N_2$-$CH_4$ mixture with plasma off (red) and plasma on (black). The consumption of methane at m/z 16 and its fragments m/z 15, 13, 12 is visible after the plasma is switched on. (b) Lower panel: With cold trap. Mass spectra of a 90-10% $N_2$-$CH_4$ mixture with plasma off (red) and plasma on (black). The main products seen in the top plot ($C_2$, $C_3$ and $C_4$) decrease by about two orders of magnitude in intensity in the bottom figure. This shows the efficient role of the cryotrap, which is to trap (albeit not completely here) the gas products formed in the plasma.

Fig. 2 shows two sets of $N_2$-$CH_4$ mass spectra, one (in red) taken before the plasma is switched on, the other (in black) taken 5 min after plasma ignition, thus making sure methane dissociation and the plasma have reached their equilibrium state (Sciamma-O'Brien et al., 2010; Alves et al., 2012).

The m/z 29 and 30 peaks are the isotopic signatures of $N_2$. Under usual conditions, without cold-trap, several mg/h of solid particles are produced in the plasma volume in steady state conditions after the very first minutes of the discharge (Sciamma-O'Brien et al., 2010). Now, with the cold-trap, solid particles are no longer produced even after a 2h plasma duration: the aerosol formation is prevented. So, we must investigate how far the gas-phase chemistry goes in the presence of the cold trap. The plasma is out of thermodynamical equilibrium and hosts fast ion-neutral reactions. If the entrapment is faster than the chemistry, then the absence of aerosols might be explained by the fact that any product larger than $CH_4$ would be removed from the reaction space. But, if the chemical kinetics of the gas phase chemistry is competitive with the entrapment, then complex molecules are produced before their entrapment on the cold reactor walls. In this case, the absence of aerosols is caused by the low amount of gas-phase precursors in the discharge, thus preventing the gas-to-solid conversion. The analysis of the mass spectra obtained *in situ* during the plasma discharge with and without cold-trap (Fig. 2) informs us on this effect of the cold-trap on the gas-phase chemistry.

The methane consumption can be estimated with m/z 15 and m/z 16 on Fig. 2 corresponding to the $CH_3^+$ and $CH_4^+$ fragments. Following the method described in Sciamma-O'Brien et al., 2010, we quantify the methane consumption with the $CH_3^+$ (m/z 15) fragment at a pressure of 0.9 mbar. With the current cold trap setup, we found this consumption to be of 52% and 60% at 10% and 1% $[CH_4]_0$, respectively. Without the current cryotrap system, Sciamma-O'Brien et al., 2010 found methane consumption efficiencies of 45% and 82% at 10% and 1% $[CH_4]_0$. The behavior of methane consumption obtained with and without cold trap are comparable, even if an effect of the different electrode temperature cannot be discarded.

In the following sections, we define $C_n$ families as molecular classes using a $C_xH_yN_z$ nomenclature for N-bearing and hydrocarbon species, where $x + z = n$. As such, $C_n$ blocks detected using mass spectrometry can be traced, where *n* is related to the presence of heavy aliphatic and N-bearing compounds.

In the $C_1$ block, in addition to the methane ion peaks, we can notice the signature of the residual water of the mass spectrometer at m/z 17, 18. Other compounds due to the residual air in the mass spectrometer can be seen at m/z 32 ($O_2$) and m/z 44 ($CO_2$).

The role of the cryotrap has a substantive effect on trapping organic volatile products. Heavy compounds such as the $C_4$ block (Fig. 2a) vanish with the cold trap (Fig. 2b), despite similar methane consumptions.

This shows the role of the cryo-trap, which is to efficiently condense and accumulate *in situ* the organic products on the cold walls of the reactor instead of being evacuated out as volatiles during the discharge.

Methane is consumed in similar quantities in the plasma discharge with and without the cryotrap. Some large and complex molecules are still detected during the discharge by mass spectrometry (Fig. 2b), up to the upper mass limits of our instrument (m/z 100) with the cold trap, showing that the gas phase chemistry ensuring their formation is still occurring. Their concentration in the gas phase is drastically reduced, by two orders of magnitude (Fig. 2b), suggesting that they are efficiently trapped on the plasma walls while they are forming. The cold trap does not prevent the formation of complex gas-phase molecules, but mainly blocks the gas-to-solid conversion step. The cryogenic trap stops the gas-phase chemistry in the stage preceding the formation of solid organic aerosols. It condenses the gas-phase precursors while they are forming, and no aerosols are observed during the 2h plasma discharge.

### *3.1 Release of volatiles back to room temperature*

We leave the cold trap continuously on during a 2h plasma discharge. After 2h, the plasma and cryo-trap are stopped, the chamber is pumped down to $10^{-4}$ mbar, isolated, and left to go back to room temperature. Hence, all the species condensed on the cold walls are progressively desorbed according to their vaporization temperatures and analyzed *in situ*. We define MS1, MS2 and MS3 (Figures 3 and 4) as three mass spectra for 10% $[CH_4]_0$ taken at three different pressures (~ 0.38, 0.74 and 1.84 mbar) and temperatures (-130ºC, -79ºC and +22ºC) during volatile release, respectively. The temperature heat-up rate initially increases relatively fast (~ 7 ºC/min between t = 0 min and the acquisition of the MS1 spectrum), then decreases to ~ 0.3 ºC/min near MS2 and finally reaches a temperature decrease rate of ~ 0.1 ºC/min above -50ºC (MS3).

This desorption is recorded with a pressure gauge within the isolated chamber as the temperature increases. Figure 3 (top) shows the pressure of the gas products being released and the temperature of the electrode (bottom) according to time, both at $[CH_4]_0$ = 10% and $[CH_4]_0$ = 1%, in blue and red, respectively. Both pressures increase, owing to the volatiles sublimating though in different behaviors. Most of the volatiles have been released at ~6h of release (Figure 3, top) and the curve starts to reach an asymptotic limit in the case of 10% methane, (i) the release starts immediately with the temperature increase until 0.74 mbar at -79ºC where it reaches a small plateau, followed by (ii) a second volatile release starting at -75ºC with a sharp decrease in heating rate at 300 min until it (iii) reaches a second plateau at 1.84 mbar. On the other hand, products formed in $[CH_4]_0$ = 1% conditions have a different release pattern. Notwithstanding the discrepancy in final pressure obtained (0.35 mbar, 5x less), the products take a much longer time to be desorbed, the release starting at about 90 min with a much smoother slope. Unlike at 10% methane, there is no plateau; the release of the volatiles is continuous and in a rather limited pressure range.

These qualitative differences suggest, as a first approximation, different products in the two methane conditions. In particular, all the products released at temperatures below -100ºC at 10% methane are absent in the case of 1% methane. This means that these specific compounds are not produced in significant amounts in the plasma discharge at 1% $CH_4$.

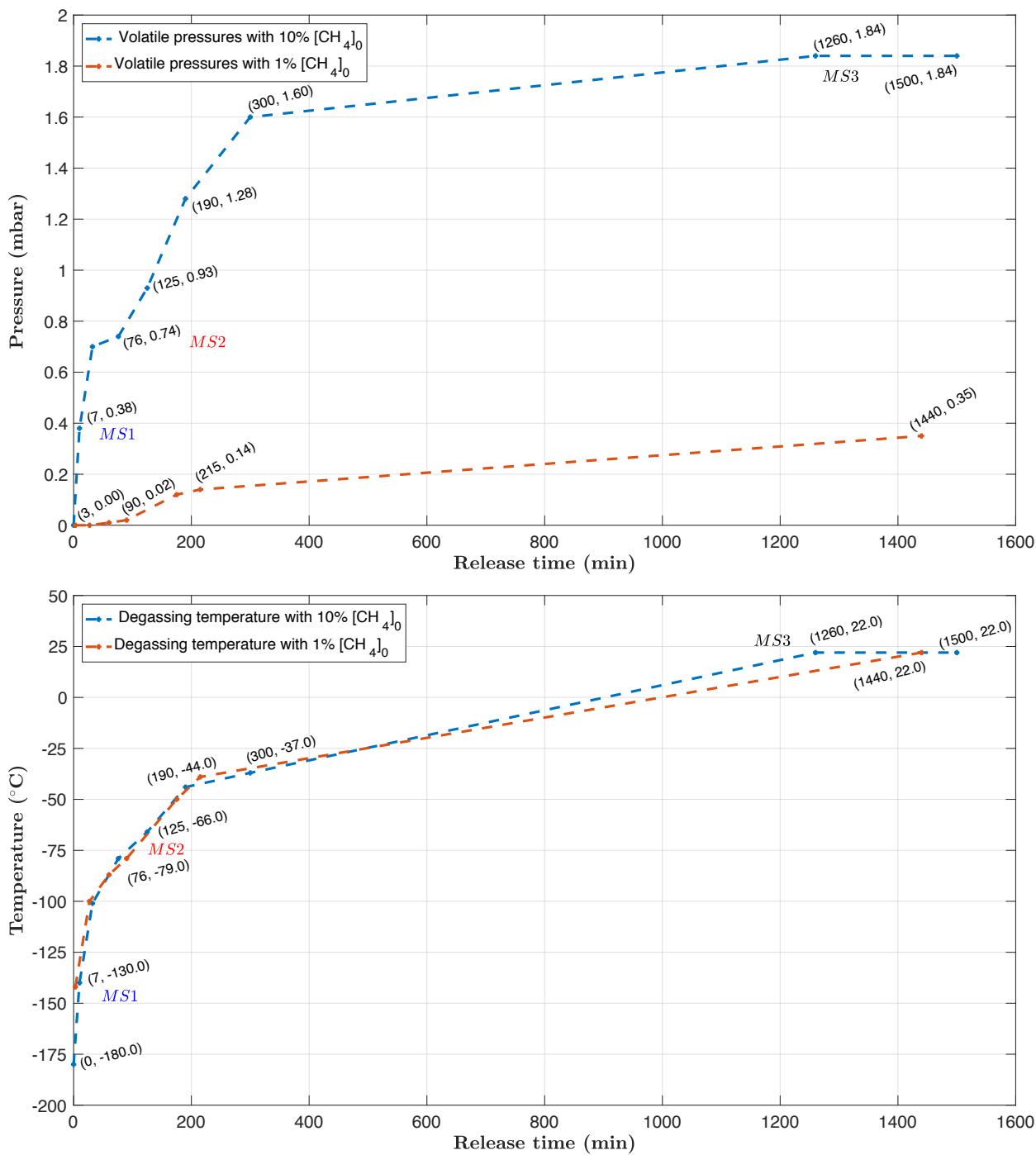

**Fig. 3**. (Top) Pressure evolution of the gas products according to degassing time in a 90-10% (blue) and 99-1% $N_2$-$CH_4$ (red) gas mixture. Labeled at each data point are the times and vessel pressures, respectively. The *MS1* (-130ºC, 0.38 mbar), *MS2* (-79ºC, 0.74 mbar), *MS3* (+22ºC, 1.84 mbar) color-coded labels correspond to the three spectra of Fig. 4 taken at their corresponding temperatures and pressures, underlying significant detections at 10% methane. (Bottom) Temperature evolution during degassing of the volatiles with the same color code for both mixing ratios.

### 3.1.1. Products released at temperatures below -100°C in the case of [CH$_4$]$_0$ = 10%,

We are first interested in the molecules released below -100°C in the case of [CH$_4$]$_0$ = 10%, which are not produced (or at least not in detectable quantities) in [CH$_4$]$_0$ = 1%.

As a first qualitative guess, we use the Antoine Law (Mahjoub et al., 2014) Eq. (1) to calculate the vapor pressure of C$_2$H$_6$, which, at -180°C, is slightly above the triple point of -183.15°C (Fray and Schmitt, 2009). C$_2$H$_6$ is indeed one of four volatiles expected to be present our final gas mixture (Gautier et al., 2011), along with C$_2$H$_2$, C$_2$H$_4$, and HCN. We use the A, B, C parameters from the NIST database (Carruth and Kobayashi, 1973). The low temperature used in the present study is at the edge of the validity range for C$_2$H$_6$. Nonetheless, we extend this law's validity domain and assume it to be applicable to these low-temperature conditions and thus must consider this calculation as a first approximation only. The vapor pressure is calculated at -180°C.

$$\log_{10}(P_s) = A - \left(\frac{B}{T+C}\right) \qquad (1)$$

For C$_2$H$_6$, we find P$_s$ = 5.7 x 10$^2$ mbar. This vapor pressure combined with the fact that we are slightly above the triple point of C$_2$H$_6$ near the liquid/vapor equilibrium, shows that this species will hardly condense.

HCN, C$_2$H$_2$ and C$_2$H$_4$ all have their triple points (Fray and Schmitt, 2009) at temperatures higher than -180°C. They are listed in Table 1. Fray and Schmitt, 2009 provided a review of vapor pressure relations obtained empirically and theoretically for a wide variety of astrophysical ices. They fitted polynomial expressions to extrapolate the sublimation pressures and used theoretical and empirical interpolation relations to obtain the coefficients of the polynomials. The polynomial expression from Fray and Schmitt, 2009 is given Equ. 2. The polynomials (Table 1) are noted $A_i$ and $T_p$ are the corresponding temperatures, -180°C (93K) in our case. The sublimation pressures (P$_{sub}$) are presented in Table 1.

$$\ln(P_{sub}) = A_0 + \sum_{i=1}^{n} A_i / T_i \qquad (2)$$

| | $T_p$ (°C) | $A_0$ | $A_1$ | $A_2$ | $A_3$ | $A_4$ | $P_{Sub}$ (-180°C) in mbar |
|---|---|---|---|---|---|---|---|
| HCN | -14.15 | $1.39 \times 10^1$ | $-3.62 \times 10^3$ | $-1.33 \times 10^5$ | $6.31 \times 10^6$ | $-1.13 \times 10^8$ | $1.69 \times 10^{-12}$ |
| $C_2H_2$ | -81.15 | $1.34 \times 10^1$ | $-2.54 \times 10^3$ | 0 | 0 | 0 | $9.48 \times 10^{-4}$ |
| $C_2H_4$ | -169.15 | $1.54 \times 10^1$ | $-2.21 \times 10^3$ | $-1.22 \times 10^4$ | $2.84 \times 10^5$ | $-2.20 \times 10^6$ | $1.3 \times 10^3$ |

**Table 1.** Vapor pressures $P_{sub}$ (mbar) calculated at 93K (-180°C) for HCN, $C_2H_2$ and $C_2H_4$. The $T_p$ and $A_i$ columns correspond to the triple points and coefficients of the polynomials of extrapolations, respectively, as given by Fray and Schmitt, 2009.

The larger values (Table 1) for the two $C_2$ hydrocarbons in comparison with HCN agree with a first substantial release of the light $C_2$ hydrocarbons at temperatures below -100°C. HCN contributes to the gas composition mainly in the second stage, for temperatures larger than -80°C, because of its low vapor pressure at colder temperatures.

The absence of $C_2$ hydrocarbons in the case of $[CH_4]_0$ = 1% is in agreement with the *ex-situ* analysis made in Gautier et al., (2011).

*3.1.2. Detections by mass spectrometry*

To go further in the analysis of the gas release, we monitored the gas phase composition by mass spectrometry at three threshold moments *MS1*, *MS2*, *MS3* for 10% methane and only in the final gas mixture for 1%.

In the case of 10% methane (Fig. 3 dashed blue line), the assigned *MS1*, *MS2* and *MS3* spectra are defined as the T, P conditions of -130°C and 0.38 mbar, -79°C and 0.74 mbar and +22°C and 1.84 mbar of products formed, respectively. *MS1* is representative of the first release stage occurring at temperatures lower than -100°C, *MS2* is on the pressure plateau beginning at -100°C, and *MS3* corresponds to the final gas mixture obtained. For 1% methane conditions, we will only focus on the last point, where a substantial volatile pressure was measured.

A first iteration at $[CH_4]_0$ = 10% was done. Mass spectra of the volatiles are shown Fig. 4. at different stages in their release after the plasma discharge, namely *MS1*, *MS2* and *MS3*.

- *MS1*: 7min after stopping the cryogenic trap (blue line), $C_1$, $C_2$ and $C_3$ block species have started to appear, with $C_1$ and $C_2$ (light volatiles of m/z < 30 amu) dominating the spectrum. Masses 15 and 28 stand out for the $C_1$ and $C_2$ blocks, respectively. Mass 28 cannot be attributed to $N_2$ as it was not trapped at -180°C during the experiment. It can be attributed to ethylene $C_2H_4$ or ethane $C_2H_6$ as both molecules are expected to be released at temperatures lower than -100°C and both have their major fragment at m/z 28 according to the NIST database. However, the previous ethane vapor pressure calculations and its triple point (Fray and Schmitt, 2009) show that it could not condense. In these plasma conditions, it is risky to attribute m/z 29 to methanimine ($CH_2=NH$) as the contribution of propane ($C_3H_8$) becomes non negligible (Carrasco et al., 2012). m/z 29 corresponds indeed to the main peak in the fragmentation pattern of $C_3H_8$. A few heavier volatiles in the $C_3$ and $C_4$ block are also beginning to be detected. At m/z 44 we could also suspect the contribution of $CO_2$. A blank mass spectrum of the mass spectrometer only has previously been measured (Carrasco et al., 2012), showing the same intensity of $2 \times 10^{-12}$ A at m/z 44 as the one seen in Fig. 4 (blue plot), at low temperature. It corresponds to the weak air signature within the mass spectrometer at a vacuum limit of about $3 \times 10^{-8}$ mbar. Consequently, the contribution of $CO_2$ observed in the mass spectra of Fig. 4 at m/z 44 is minor inside the chamber among the released products and corresponds to the residual air in the mass spectrometer itself.

- *MS2*: The main differences compared to *MS1* revolve around the $C_3$ and $C_4$ blocks with the appearance of heavier hydrocarbons (red line, m/z > 50). Within the $C_3$ block, an important contribution appears at m/z 39, consistent with the cyclopropene isomer $C_3H_4$. A significant increase is also observed at m/z 40 and 44 compatible with the fragmentation pattern of propane $C_3H_8$ in the NIST database. No species with masses higher than m/z 58 are detected yet.

- *MS3*: Remarkable changes occur at this final stage (black line), where many new species stand out. In the $C_1$ block, m/z 17 ($NH_3$) strongly increases to become the dominating $C_1$ compound.

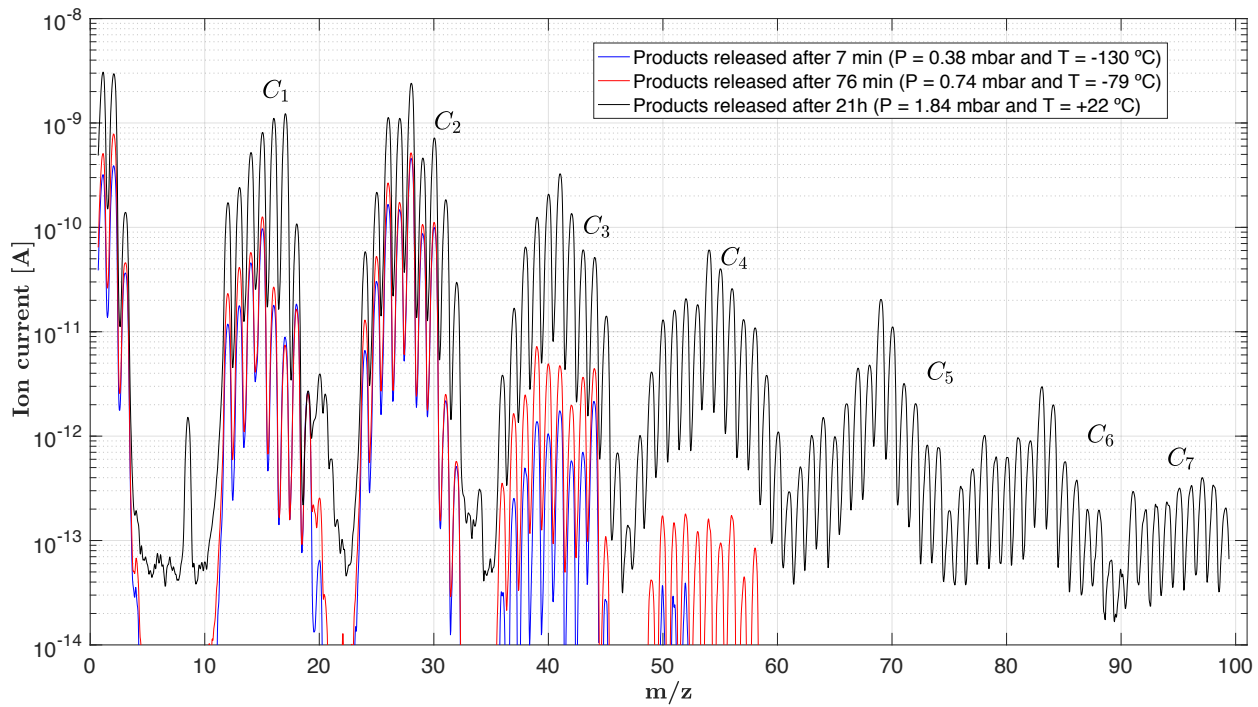

**Fig. 4.** Detection and evolution of several $C_n$ blocks for $[CH_4]_0$ = 10%. Plots marked in blue, red and black represent intermittent spectra taken 7 min, 76 min and 21h after commencing volatile release back to room temperature (*MS1*, *MS2* and *MS3*, respectively). The indicated temperatures correspond to those taken at the start of a mass scan. Note that during a scan acquisition (~200s long), the temperature may change over 0.3 ºC to 23 ºC depending on the heating rates. The color code used here is the same as *MS1*, *MS2* and *MS3* of Fig. 3.

If we now compare the final stages obtained in the case of $[CH_4]_0$ = 1% and 10% on Fig. 3, the total pressure of volatiles obtained (0.35 mbar, blue line) is much less than in the case of $[CH_4]_0$ = 10% (red). Fig. 5 shows a comparison of the final states at $[CH_4]_0$ = 1% (brown) and $[CH_4]_0$ = 10% (blue), with the initial state (black), taken just before the volatile release at low-controlled temperature T = -180ºC. As expected from the lower final pressure, the mass spectrum is much less intense in the case of $[CH_4]_0$ = 1% than 10%. No signature is detected beyond the $C_5$ block for 1% whereas large intensities are found until the $C_7$ block in the case of 10%. However, the major peaks are the same in both cases. The $C_1$ block is dominated by m/z 17, accountable for ammonia. This important production, discussed in section 4.2, is a new result compared to the work by Gautier et al. (2011) and Carrasco et al. (2012). Indeed, as detailed in Carrasco et al. (2012), ammonia was hardly detectable with the analytical methods used in the two latter studies. The $C_2$ block has high peak intensities at m/z 26, 27, 28, 30 for $[CH_4]_0$ = 10% but the one at 28 stands out in both methane conditions, compatible with $C_2H_4$ or $C_2H_6$. The $C_3$ block is dominated by m/z 41, accountable for acetonitrile $CH_3CN$; m/z 54 dominates $C_4$, which could correspond to butadiene $C_4H_6$ or $C_2N_2H_2$ (HCN dimer). For $[CH_4]_0$ = 10%, the $C_5$, $C_6$ and $C_7$ blocks are dominated by masses 69, 83 and 97, respectively. The odd masses predominantly detected suggest a strong contribution of N-bearing molecules in both conditions.

The intensity evolution of m/z 17, 26, 27, 28 and 41 over time are shown Figure 11 in the Supplementary Material.

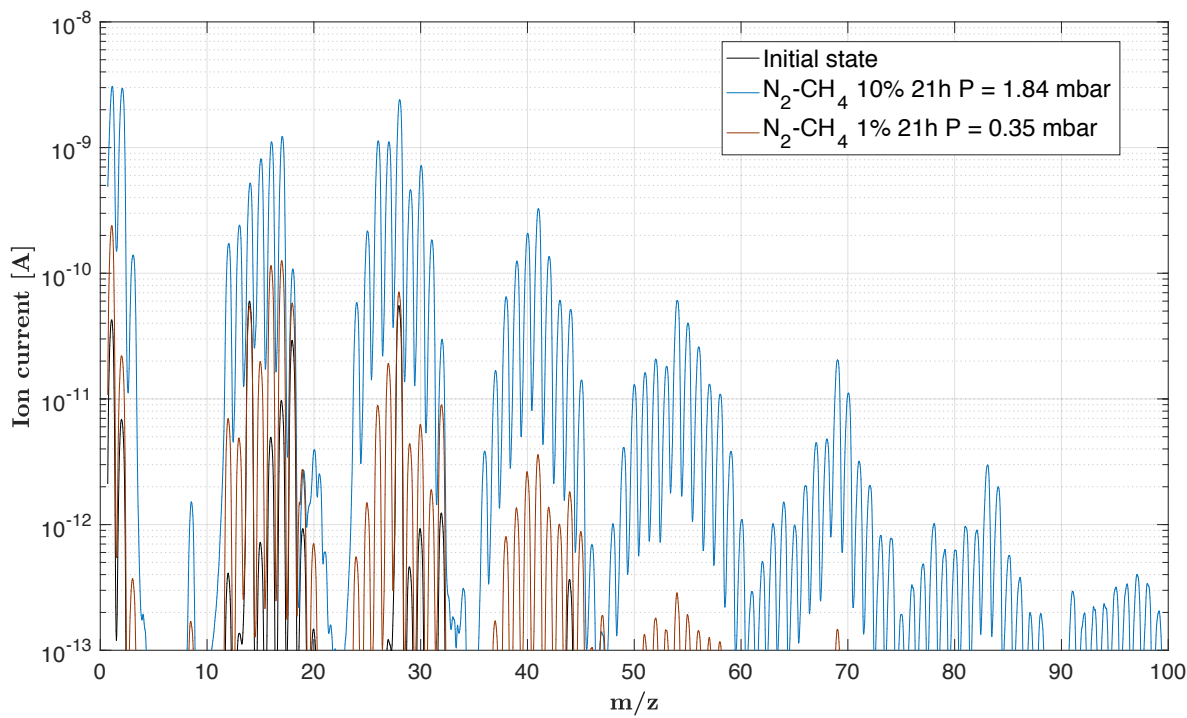

**Fig. 5.** Three superimposed spectra at different $[CH_4]_0$ concentrations. In black, the initial mass spectrum taken before release of the volatiles, still at low-controlled temperature (and representative of the blank of our mass spectrometer). In blue and brown, the final state of volatiles at $[CH_4]_0$ = 10% after 21h and $[CH_4]_0$ = 1% after 21h of release, respectively.

### *3.2 Monitoring and quantification using Mid-Infrared spectroscopy*

Alongside the neutral mass analysis, infrared spectra are simultaneously taken throughout the release of the volatiles inside the chamber to provide quantification of these volatiles.

The measurement range is 650 $cm^{-1}$ to 4000 $cm^{-1}$ (2.5 μm to 15.4 μm) at a resolution of 1 $cm^{-1}$ during the release of the gas phase products analyzed *in situ*. Fig. 6 shows the spectra taken in both initial methane conditions. The top plot was taken for a $[CH_4]_0$ = 10%, the bottom at $[CH_4]_0$ = 1%.

There is a clear discrepancy in the volatile products formed between the two conditions. $[CH_4]_0$ = 10% produces –CH and –CH$_2$ compounds visible in the stretching mode region 3,000-3,500 cm$^{-1}$ that are absent in the $[CH_4]_0$ = 1% case. Likewise, there is a relatively far greater absorption of molecules bearing C=N and C=C bonds (~1,500 cm$^{-1}$) with $[CH_4]_0$ = 10%. Fig. 6 (bottom) also has a large, broad absorption spread out over 2,000-2,500 cm$^{-1}$ absent in the top figure. These signatures could correspond to a scattering feature of solid grains in suspension which could be the result of the released precursors reacting post-discharge. Furthermore, in Sciamma-O'Brien et al. (2010), it had been shown that the gas to solid conversion yield was more important in the case of the 1% experiment than in the 10% experiment, in agreement with a higher amount of N-bearing gas phase products compared to hydrocarbons in the 1% experiment. The present results are consistent with this previous study.

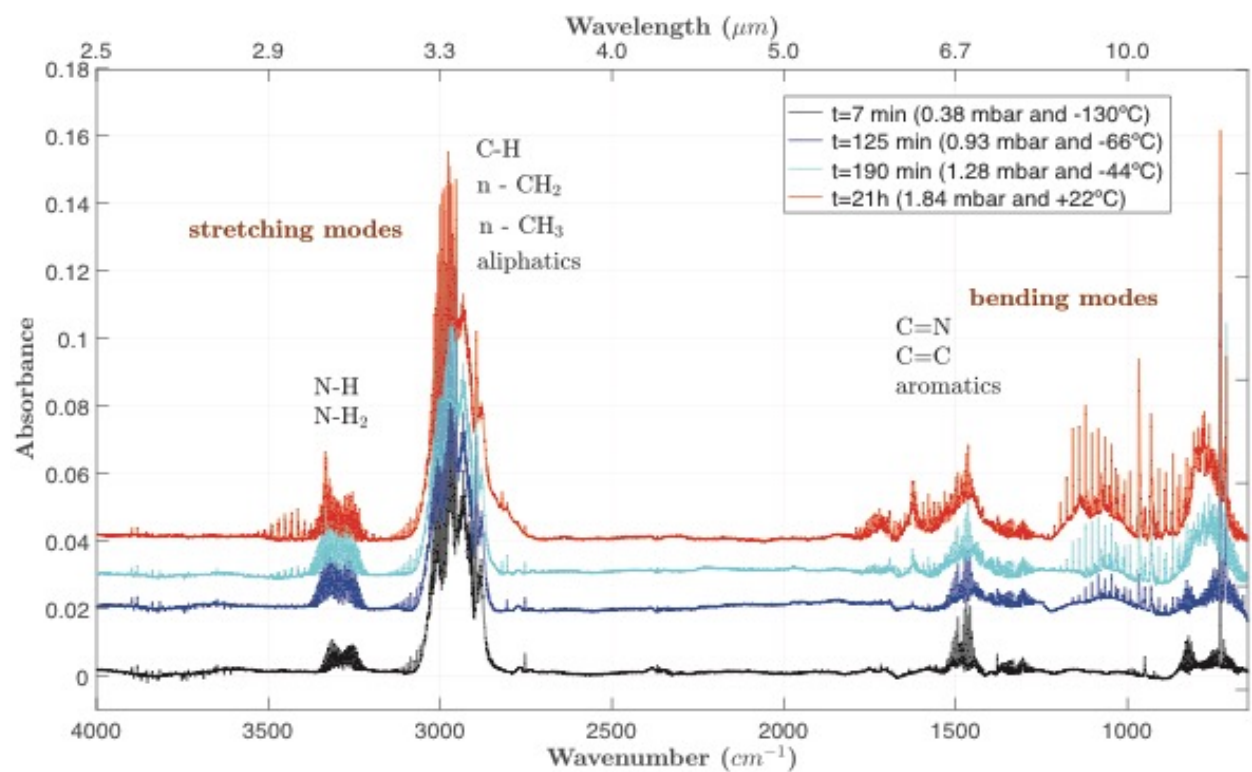
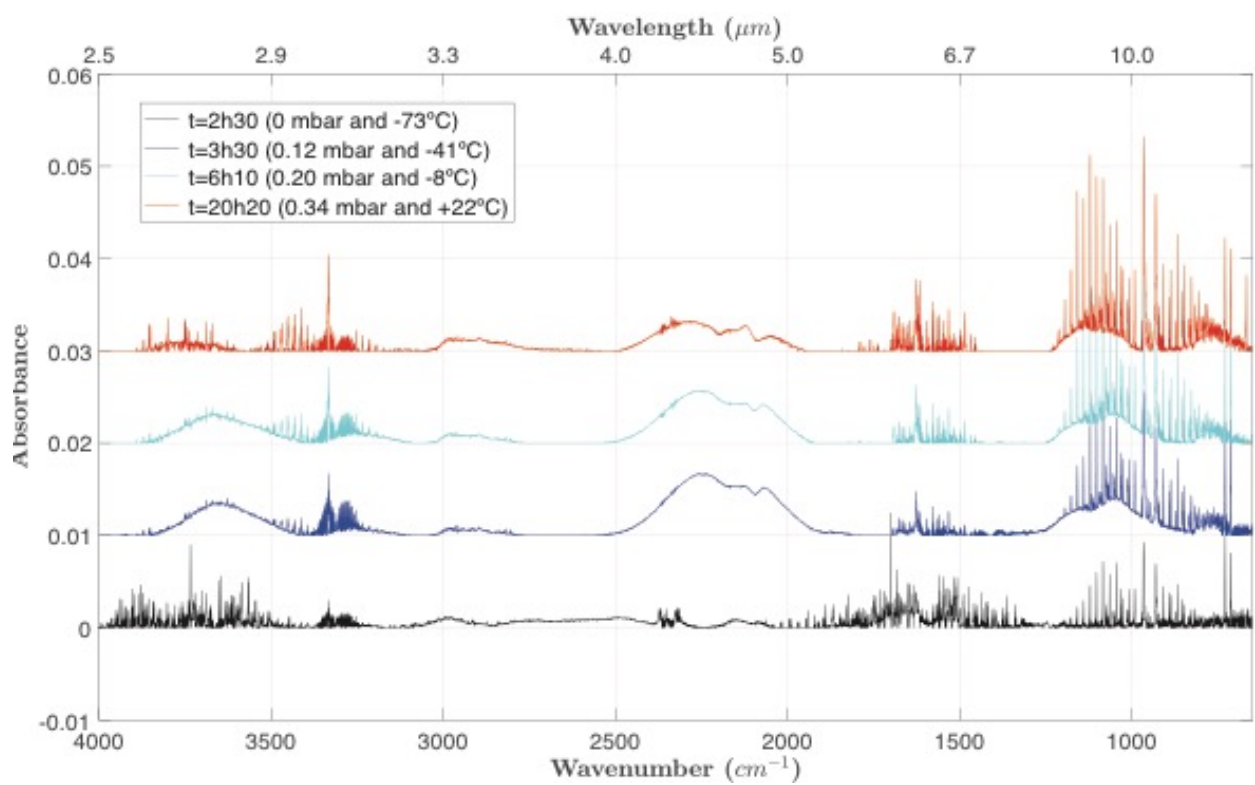

**Fig. 6** FT-IR spectra of the volatiles taken after the 90-10% (top) and 99-1% (bottom) $N_2$-$CH_4$ plasma conditions, plotted with an arbitrary absorbance against a 650-4000 $cm^{-1}$ wavenumber range. The volatile density produced during the plasma discharge is being incrementally released and analyzed through IR spectroscopy. Top: Total gas pressures of 0.38 mbar (-130ºC), 0.93 mbar (-66ºC), 1.28 mbar (-44ºC) and 1.84 mbar (+22ºC) are shown in black, blue, cyan and red, respectively. Bottom: Total measured gas pressures of 0 mbar (-73ºC), 0.12 mbar (-41ºC), 0.20 mbar (-8ºC) and 0.34 mbar (+22ºC) with the same color code. One clear difference is in the absence of any substantial aliphatic compounds (2800-3100 $cm^{-1}$) at $[CH_4]_0$ = 1% (bottom) that stand out at $[CH_4]_0$ = 10% (top).

Among all the signatures observed, we identified and calculated the concentrations of four volatiles as they are released in the chamber: $NH_3$, $C_2H_2$, $C_2H_4$, and HCN. These species are listed in Table 2. Ammonia ($NH_3$), acetylene ($C_2H_2$), hydrogen cyanide (HCN), ethylene ($C_2H_4$), are all species present at ionospheric and/or stratospheric altitudes on Titan, considered to be key in the formation of Titan's aerosols (Hanel et al., 1981; Kunde et al., 1981; Maguire et al., 1981; Wilson and Atreya, 2003, 2004) and formed in the upper atmosphere (Hörst, 2017).

| Selected species | $C_xH_yN_z$ | Main absorption bands ($cm^{-1}$) | Detected in Titan's atmosphere |
|---|---|---|---|
| Ammonia | $NH_3$ | 962 (10.4 µm) | 1, 4 (upper limit inferred) |
| Acetylene | $C_2H_2$ | 729.25 (13.7 µm) | 1, 2, 6 |
| Hydrogen cyanide | HCN | 712.3 (14.0 µm) | 1, 5, 6, 7, 8 |
| Ethylene | $C_2H_4$ | 949.3 (10.5 µm) | 1, 3 |

**Table 2.** Four major volatile compounds detected and analyzed by infrared spectroscopy at $[CH_4]_0$ = 1% and $[CH_4]_0$ = 10%. The main infrared absorption bands which were used for density calculations are also given. References are 1: Vinatier et al. (2007), 2: Cui et al. (2009), 3: Coustenis et al. (2007), 4: Nelson et al. (2009), 5: Paubert et al. (1984), 6: Teanby et al. (2007), 7: Moreno et al. (2015), 8: Molter et al. (2016). For more details on the bands and absorption cross-sections used for the molecular density calculations, the reader is referred to Table 3 and Figures 9 and 10 of the Supplementary Material.

To track the production and evolution with time/temperature of these four compounds, we used *in situ* infrared quantification. By using Beer-Lambert's Law (Eq. 3), we can calculate the concentration of any given gas phase product.

$$I_t = I_0 \times e^{-l.\sigma.N} \quad (3)$$

$I_t$, $I_0$ and $l$ are, respectively, the transmitted, incident intensities, optical path (cm) and $\sigma$ the absorption cross-section ($cm^2.molecule^{-1}$) given by the Pacific Northwest National Lab (PNNL), University of Washington (Sharpe et al., 2004), GEISA (Armante et al., 2016) and ExoMol (Harris et al., 2006; Hill et al., 2013; Tennyson et al., 2016) databases. In addition, *N* is the molecular density ($cm^{-3}$). The absorption *A* is as follows:

$$A_{(\lambda)} = l \times \sigma_{(\lambda)} \times N \quad (4)$$

Using the integrated experimental absorption *A* with the integrated database absorption cross-section $\sigma$ between the two wavelengths $\lambda_1$ and $\lambda_2$, (4) becomes:

$$\int_{\lambda_1}^{\lambda_2} A \, d\lambda = l \times N \times \int_{\lambda_1}^{\lambda_2} \sigma \, d\lambda \quad (5)$$

So, we obtain the molecular number density *N* (molecules.$cm^{-3}$):

$$N = \frac{1}{l} \times \frac{\int_{\lambda_1}^{\lambda_2} A \, d\lambda}{\int_{\lambda_1}^{\lambda_2} \sigma \, d\lambda} \quad (6)$$

Number densities are derived over a wavenumber range (see Table 3 of the Supplementary Material), covering the absorption band considered for each species.

The main absorption band of each species was taken for all cases. The density results for both conditions are shown Fig. 7 and Fig. 8. For the detailed number density calculation results and data points used with the specific integration bands, the reader is invited to peruse the Supplementary Material of the online version of this article.

The 962 cm$^{-1}$ (10.4 µm) $NH_3$ assigned band refers to the $\upsilon_2$ N-H symmetric deformation. It is part of the $NH_3$ doublet at 930 and 960 cm$^{-1}$, caused by the motion of nitrogen through the plane of the three protons leading to an energy barrier. In the case of $C_2H_2$, we used acetylene's 729.25 cm$^{-1}$ frequency of oscillation (13.7 µm). This band corresponds to the symmetric CH bending mode $\upsilon_5$ of every other C and H atom to its respective neighboring atom. For HCN, we used the fundamental $\upsilon_2$ bend frequency at 712 cm$^{-1}$ (14.0 µm); $C_2H_4$ the $\upsilon_7$ $CH_2$ wag deformation at 949 cm$^{-1}$ (10.5 µm). $C_2H_4$ centered at 950 cm$^{-1}$ is surrounded by the $NH_3$ doublet at 930 and 960 cm$^{-1}$. This absorption represents the fundamental absorption of ammonia's $\upsilon_2$ normal vibrational mode. The motion of nitrogen through the plane of the three protons causes this vibration with an energy barrier. For a summary on each band and frequency used, the reader is referred to Table 3 of the Supplementary Material.

Molecular number densities at $N_2/CH_4$ : 99/1% are listed in Table T-4. The measurements taken at $t_{1-6}$ correspond approximately to 30 min, 60 min, 120 min, 180 min, 1200 min and 1440 min, respectively. Three datasets were acquired after three experiments in the same initial conditions. At this point, it is noteworthy to remember that $t_0$ represents the time when the plasma conditions and cryo-trap cooling are ended (Fig. 3), and thus when the condensed volatiles can effectively desorb from the chamber still under vacuum walls into the gas phase.

Overall, ammonia is the most dominant $C_1$ neutral (~ 1.1 x 10$^{15}$ cm$^{-3}$) produced at 1% $CH_4$, confirming its high intensity at m/z 17 detected in mass spectrometry (Fig. 4). $C_2H_2$ and $C_2H_4$ also reached relatively high abundance ~ 3.2 x 10$^{12}$ cm$^{-3}$ and ~ 9.9 x 10$^{12}$ cm$^{-3}$, respectively. These number densities correspond to the final measurements.

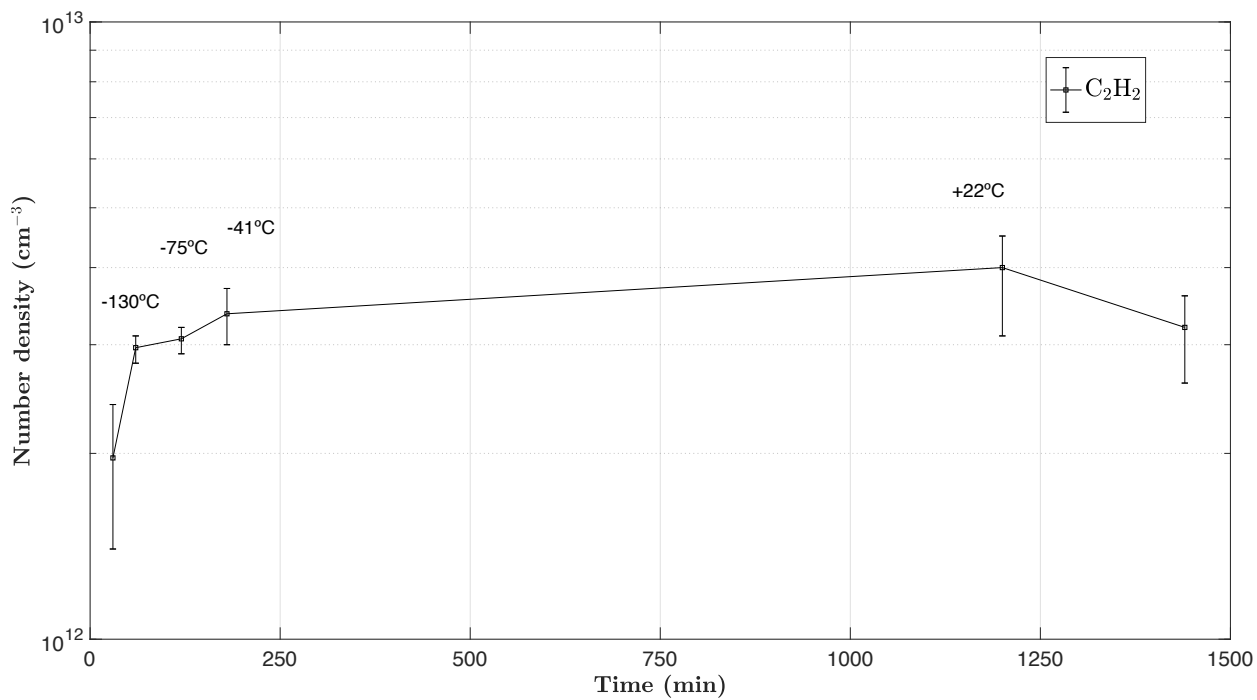

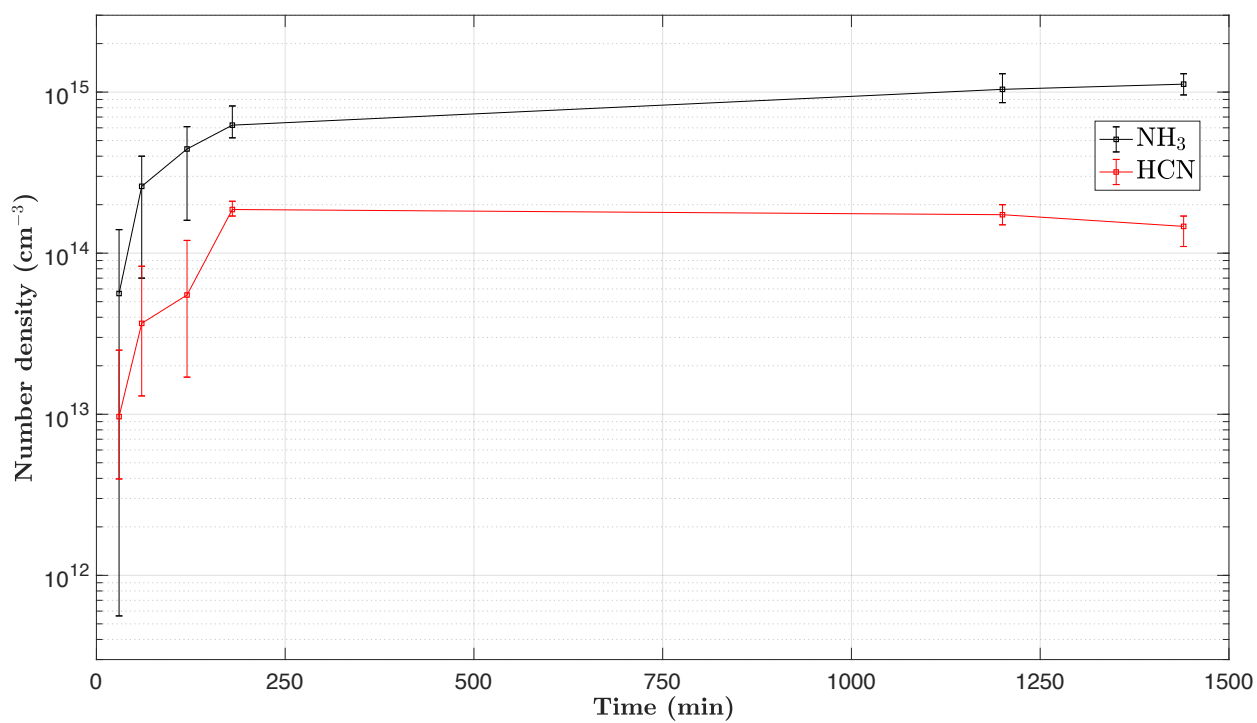

**Fig. 7.** 1% methane conditions. Top: Molecular densities and average fitted kinetic profiles of $C_2H_2$ and $C_2H_4$. Bottom: $NH_3$ and HCN. Approximate temperatures are also labeled over each data point. The error bars represent the dispersion of the data points for all three experiments (Table 4 and 5 of the online version of this article).

Fig. 8 shows the same kinetic profiles from a 10% methane initial gas mixture, with the molecular densities listed in Table T-4. The $C_2$ species all reach concentrations at an order of magnitude higher than those at 1%, correlating with the higher methane concentration initially injected, which promotes the production of hydrocarbon species. $C_2$ species are thus strongly favorably produced with increasing methane concentrations. This is especially notable with the amount of HCN produced (~ $1.2 \times 10^{15}$ cm$^{-3}$) in the 10% $[CH_4]_0$ condition.

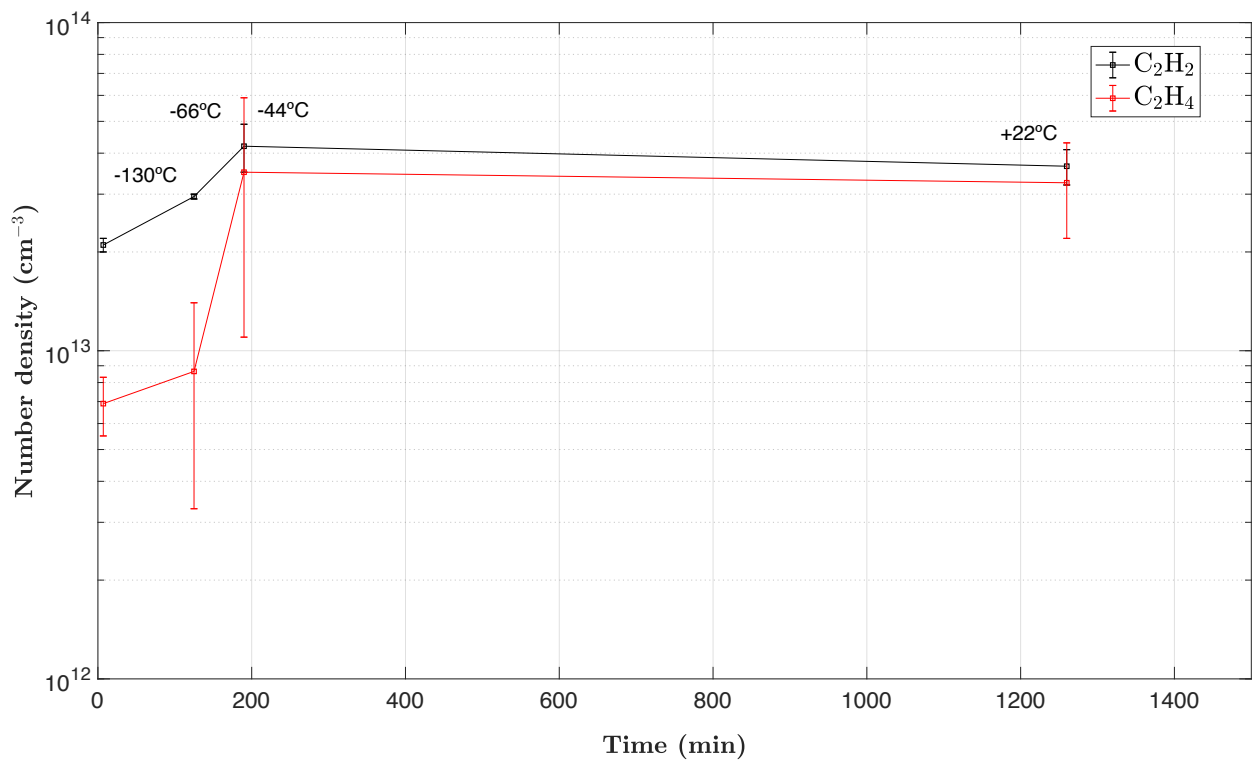

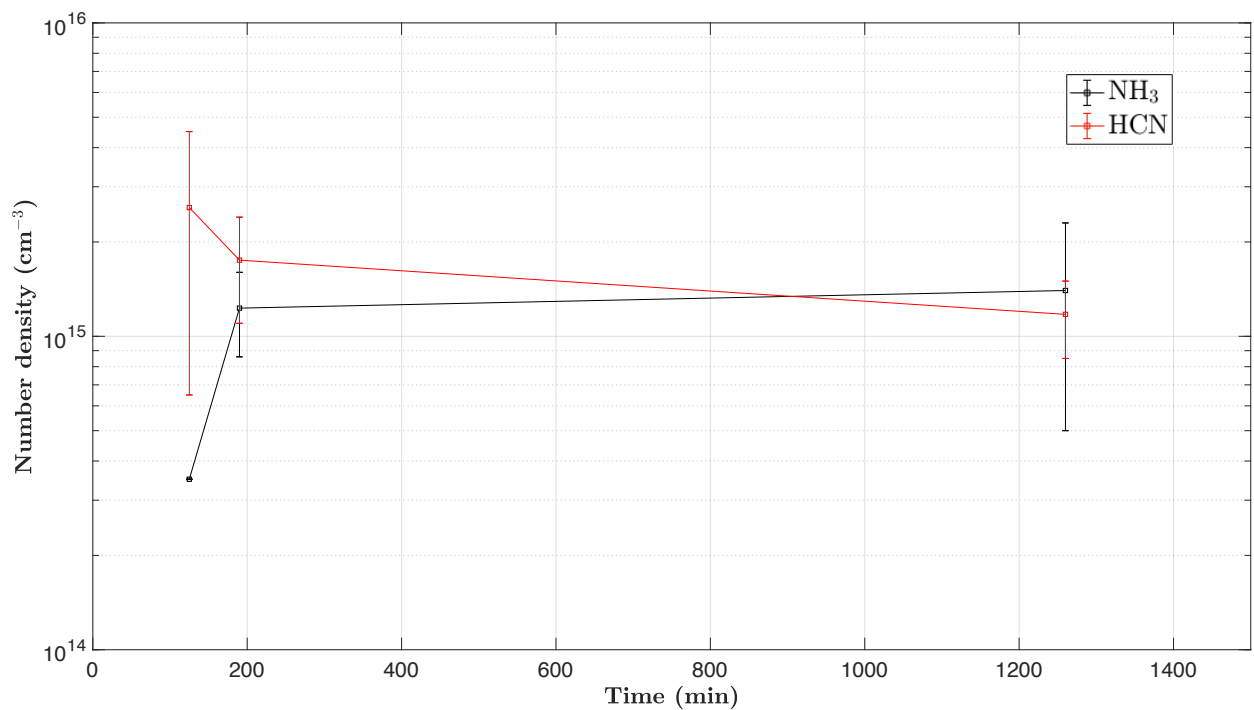

**Fig. 8.** 10% methane conditions. Top: Molecular densities and average fitted kinetic profiles of $C_2H_2$ and $C_2H_4$. Bottom: $NH_3$ and HCN. Approximate temperatures are also labeled over each data point. The error bars represent the dispersion of the data points for both experiments (Table 4 and 5 of the online version of this article).

### 3.3 Methane consumption and hydrocarbon yield

With a 10% CH$_4$ mixing ratio, the condition with the most products, we consider the CH$_4$ consumption to provide yield estimates on the carbon conversion to hydrocarbon gas products. We note *D* as the total gas flow rate (N$_2$-CH$_4$) into the chamber in standard conditions. We first calculate the amount of CH$_4$ consumed in standard conditions (mg.h$^{-1}$), with *D* = 55 sccm. If $n_{tot}$ is the total number of gas phase moles entering the reactor per second, the ideal gas law becomes:

$$\frac{d_{ntot}}{dt} = \frac{P_0}{R \times T_0} \times D \qquad (7)$$

With $P_0$ the gas pressure in *Pa*, *R* the ideal gas constant, *T* the absolute temperature (300 K), $d_{ntot}/dt = 2.22 \times 10^{19}$ molecules.s$^{-1}$. As seen previously, with a mixing ratio of [CH$_4$]$_0$ = 10%, we find a CH$_4$ consumption of 52%. So,

$$\frac{dn_{CH_4cons}}{dt} = D_{CH_4} \times CH_{4cons} \times d_{ntot}/dt \qquad (8)$$

And the methane consumption rate is $dn_{CH_4cons}/dt = 1.15 \times 10^{18}$ molecules.s$^{-1}$. As the plasma runs for 2h, the total methane consumption is $n_{CH_4cons}(2h)/dt = 8.28 \times 10^{21}$ molecules. The total volume of the PAMPRE reactor is ~ 28×10$^3$ cm$^3$. Consequently, there is a CH$_4$ total consumption of about 2.9 × 10$^{17}$ molecules.cm$^{-3}$ over the 2h of the plasma discharge. With a calculated HCN total number of molecules of ~ 1.2 x 10$^{15}$ cm$^{-3}$, HCN accounts for ~ 0.4% of the carbon conversion from the consumed CH$_4$, while it is ~ 0.3 ‰ and 0.2 ‰ for C$_2$H$_2$ and C$_2$H$_4$, respectively.

## IV) Discussion

*4.1 Volatile discrepancies at $[CH_4]_0$ = 1% and $[CH_4]_0$ = 10%*

There are evident differences in the volatile products formed in both methane concentrations used in this study (Fig. 3, Fig. 6, Tables T-3 and T-4 of the online version of this article). Despite obtaining ~ 1.84 mbar solely in volatile products with 10% methane and ~ 0.35 mbar at 1% methane, both spectra on Fig. 5 appear to qualitatively show the same nitrogen-bearing odd mass compounds, albeit at different intensities. One major discrepancy though, lies in the overwhelming amount of hydrocarbon species produced at 10% methane. The $C_2$ block is dominated by the m/z 28 amu signal (Fig. 5), dominating the rest of the spectra, which according to the NIST database, corresponds to ethylene and ethane's main fragment (this mass cannot be linked to a release of $N_2$, as the cryotrap temperatures were not cold enough to trap any nitrogen during the experiment). However, the vapor pressure indicates that ethane is difficult to trap, and so m/z 28 can be mainly attributed to $C_2H_4$. Furthermore, the $I_{30}/I_{28}$ ratio of the absolute intensities in the mass spectra at m/z 30 and m/z 28 amu correspond to the ones given by NIST, ~ 30-35%. On Titan, short-lived species such as $C_2H_4$ remain tantalizing to study, as they experience stratospheric dynamical processes with changing mixing ratios. By preventing the condensation of $N_2$, this study shows how m/z 28 may be attributed to $C_2H_4$ as one of the most abundant hydrocarbons at 10% $CH_4$.

*4.2 Ammonia*

Moreover, $NH_3$ stands out as being the most dominant N-bearing molecule in both $[CH_4]_0$ = 1% and $[CH_4]_0$ = 10% conditions, with final calculated concentrations of ~ 1.1 x $10^{15}$ and ~ 1.4 x $10^{15}$ molecules.cm$^{-3}$, respectively.
Ammonia production was previously shown to be positively correlated with an increasing methane concentration (Carrasco et al., 2012). However, the latter study did not consider ammonia detection due to lack of a quantification approach as well as the m/z 17 $H_2O$ fragment contamination. $NH_3$, as well as HCN, are nitrogen-bearing volatiles of utmost importance related to prebiotic chemistry in reducing atmospheric environments. So, studying $NH_3$ in Titan's upper atmosphere is crucial to understanding the chemical pathways leading to the formation of aerosols and how it is incorporated in them. Cassini/INMS analyses of early Titan flybys (Vuitton et al., 2007) first reported the presence of ammonia in the ionosphere, which was later more robustly confirmed by Cui et al. (2009). Yelle et al. (2010), Carrasco et al. (2012), Loison et al. (2015) discussed $NH_3$ chemical pathways. As such, the formation of ammonia in Titan's atmosphere is still unclear, though it is thought to most likely involve methanimine $CH_2NH$ which itself acts as an important source for $NH_2$ radicals in plasma conditions, to eventually forming $NH_3$. An ionic characterization of pathways and reactions pertaining to ammonia production is out of the scope of this paper.

Yelle et al. (2010) proposed an explanation for the production of $NH_3$. Indeed, $NH_3$ is being photochemically formed in the upper atmosphere and fed down into the stratosphere. While photochemical models (Lara et al., 1996; Krasnopolsky, 2009) did not initially predict the formation of $NH_3$ in the upper atmosphere, its production can be explained through ion reactions (Yelle et al., 2010), by measuring the protonated ion $NH_4^+$. Based on the long chain reaction forming $NH_3$ detailed by Yelle et al. (2010) and its importance in our experimental results (major nitrogen-bearing molecule), the contribution of nitrogen-bearing functional groups (e.g. amines, nitriles or imine) or their contribution to eventually be incorporated into the organic aerosols seems pertinent. This appears to be the case for $NH_3$, being the main N-bearing compound. In more advanced environments whether on Titan (Neish et al., 2009) or in the interstellar medium (e.g. Largo et al., 2010), $NH_3$ remains an open and current topic to study prebiotic chemistry.

*4.3 $C_2H_4$ pathways to tholin formation*

The presence of ethylene in the ionosphere (Waite et al., 2007) and stratosphere (e.g. Coustenis et al., 2007) is well established. However, similar to the other volatiles chosen in this study, its influence and participation in tholin formation needs to be further investigated. The relative concentration of $C_2H_4$ in Titan's atmospheric column is subject to distinct features. It was first noted by Coustenis et al. (2007) and Vinatier et al. (2007) how the retrieved stratospheric vertical abundances of $C_2H_4$ from Cassini/CIRS data showed both a unique decrease in the mixing ratios towards high northern latitudes as well as being the only species whose mixing ratio decreases with altitude at 15°S near the equator. Compared to other $C_2$ molecules (e.g. $C_2H_2$, $C_2H_6$, HCN), $C_2H_4$ is unique in that its mixing ratio decreases with altitude at equatorial and northern polar latitudes. One hypothesis given by Vinatier et al. (2007) was that $C_2H_4$ (along with $C_4H_2$ and $CH_3C_2H$), is subject to a photodissociative sink along with equatorward transport from the winter polar vortex in the lower stratosphere and troposphere (Crespin et al., 2005), essentially removing it away from the gaseous atmosphere. This sink could eventually be the multiple haze layer.

As discussed previously and shown Fig. 5 and Table 5 (Supplementary Material), $C_2H_4$ is an important $C_2$ compound in the 10% $[CH_4]_0$ mixture. Similar experiments (Gautier et al., 2014) argued that (although in the absence of a cooling plasma system) tholins are formed with $CH_2$ patterns, coherent with $C_2H_4$ putatively acting as a strong gaseous precursor. This study comes in agreement, at least on a tholin formation scale that $C_2H_4$ participates in their formation. This experimental result also confirms that, if the stratospheric multiple haze layer is indeed the sink to this molecule (Vinatier et al., 2007), the latter can easily take chemical pathways leading to the formation of aerosols.

Pathways involving $C_2H_4$ (alongside $C_2H_2$) loss have been suggested by Yelle et al. (2010). They proposed that $C_2H_4$ (and $C_2H_2$) reacts with NH radicals (Reaction 9) to produce heavier nitrile molecules such as $CH_3CN$ (Reaction 10) or the cyanomethylene radical $HC_2N$.

$$N^+ + CH_4 \rightarrow CH_3^+ + NH \qquad (9)$$

$$NH + C_2H_4 \rightarrow CH_3CN + H_2 \qquad (10)$$

The formation of NH radicals, reaction (9), occurs in ion chemistry conditions through proton exchange, while these same radicals can then react with ethylene, reaction (10) in neutral conditions. Given the large abundance of produced $NH_3$, it is possible that NH radicals were formed and reacted with $C_2H_4$ during our 2h plasma, which incidentally is favorably produced with increasing $CH_4$ concentration (Reaction 9).

Interestingly, $C_2H_4$ is implicated in $NH_3$ formation. According to Yelle et al. (2010),

$$NH_2 + H_2CN \rightarrow NH_3 + HCN \qquad (11)$$

Prior to this, the formation of the amino radical $NH_2$ involves ion chemistry, with the following ion-neutral reaction:

$$N^+ + C_2H_4 \rightarrow NH_2 + C_2H_2^+ \qquad (12)$$

Thus, $C_2H_4$ in our plasma may participate in the production of heavier nitrile species or ammonia during the plasma discharge.

*4.4 HCN production*

Hydrogen cyanide in Titan's atmosphere is well documented and known to be the most abundant nitrile trace volatile (e.g. Kim et al., 2005; Vinatier et al., 2007), with high mixing ratios at stratospheric altitudes in the northern polar regions.

HCN is also of prime importance in the search for prebiotic conditions in the solar system and beyond (Oró, 1961) as well as being one of the precursors to amino acids and peptides (Oró, 1961; Hörst et al., 2012; He and Smith, 2014; Rahm et al., 2016). In addition, the chemistry of HCN mainly relies on its relatively high polarity and quite strong C≡N triple bond, with a bond dissociation energy D(H-CN) of ~ 5.2 eV calculated by photodissociation and photoionization models (Berkowitz, 1962; Davis and Okabe, 1968; Cicerone and Zellner, 1983). Thus, making it a relatively stable molecule.

As shown Figs. 7 and 8, HCN is one of the most abundant products formed in both $[CH_4]_0$ = 1% and $[CH_4]_0$ = 10% conditions, with ~ 1.5 x $10^{14}$ $cm^{-3}$ and 1.2 x $10^{15}$ $cm^{-3}$ on average, respectively. Carbon yield calculations stresses the carbon conversion to HCN (~ 0.4%), while other $C_2$ species all are favorably produced by an order of magnitude at 10% $CH_4$ than at 1% $CH_4$.

HCN and HCN-based copolymers are also known to participate in the chemical growth patterns of tholin formation (Pernot et al., 2010; Gautier et al., 2014) along with the $CH_2$ monomer, which is in agreement with the idea that haze forming acts as an HCN sink in the stratosphere (McKay, 1996; Vinatier et al., 2007). This study seems to confirm the prevalent role of HCN in the formation of tholins.

## V) Conclusions

The present study was aimed at expanding previous work focused on the volatile production and their participation in tholin production using a cold dusty plasma experiment (Gautier et al., 2011; Carrasco et al., 2012), but with one major twofold novelty (i) the ability to trap these volatiles *in situ* within our plasma reactor with a new experimental setup at low-controlled temperatures, and (ii) quantitatively analyze their formation using IR spectroscopy. It is important to be cautious in interpreting these quantitative results, as three (T, P and t) parameters change simultaneously during the sublimation phase entailing possible unsuspected ice-volatile chemistry. The approach we took was solely focused on the gas phase. Indeed, these volatiles, still poorly understood and which give Titan its unique nature of being one of the most chemically complex bodies in the Solar System, act as substantial gas phase precursors to the formation of Titan's organic aerosols populating its haze layers. Hence, simulating ionosphere conditions is crucial in constraining the volatile population, its reactivity and potential chemical pathways.

Major discrepancies exist between the two initial methane concentrations we used, 1% and 10%, with many more hydrocarbons formed in the latter case (detection of $C_7$ species), but with similar amounts of N-bearing species, only at different concentrations. Our results confirm the important role of $C_2$ species in the tholin precursor volatile family (m/z 28 e.g.), as well as that of $NH_3$. One of them, ethylene $C_2H_4$, being a relatively major $C_2$ hydrocarbons, might even reveal an important "hub" role at least at 10% $[CH_4]_0$ through which specific chemical pathways involving $C_2H_4$ are favored (see Section 4.3). The importance of $C_2H_4$ might also be relevant at 1% $[CH_4]_0$ although this is difficult to assess in this study due to the cryotrapping efficiency which is not optimum at -180ºC to quantify it, as suggested by the thermodynamical calculations and mass spectra during the discharge. Eventually, tholins themselves may be benefitting from the incorporation of $C_2H_4$, which itself stresses the value of considering a yet unaccomplished fully-coupled ion and neutral chemistry study. This is out of the scope of this paper, as the cryo-trap system prevents the formation of tholins. These experiments have also shown a strong incorporation of carbon into HCN, along with $NH_3$ being a competitive product to HCN, especially with a $[CH_4]_0$ = 1% (an order of magnitude more in final concentration).

Future compelling work to expand this study might be to consider other molecular families participating in the neutral reactivity, therefore increasing our knowledge of volatile chemistry precursor to tholin production. In addition, adding selected volatiles such as HCN, an important nitrile, in trace amounts into our gaseous mixture would improve the understanding of its influence on the chemical reactivity in plasma conditions in future work. Moreover, predicting ion densities and characterizing ion-neutral reactions in these conditions would be key to understanding their incorporation into the solid phase and validating chemical pathways which are still largely unknown in Titan ionosphere simulations. Perhaps most importantly, a fully-coupled ion-neutral characterization would fundamentally bring credence to understanding the complex organic realm encompassing Titan's gas phase chemistry.

**Acknowledgements**

We are thankful to the European Research Council Starting Grant PRIMCHEM, grant agreement no. 636829 for funding this work. We thank C. Szopa, A. Mahjoub, A. Oza and gratefully acknowledge discussions with T. Gautier. We are grateful to two anonymous reviewers for their insightful and valuable comments which improved the content of this paper.

**Supplementary Material**

**1% CH₄**

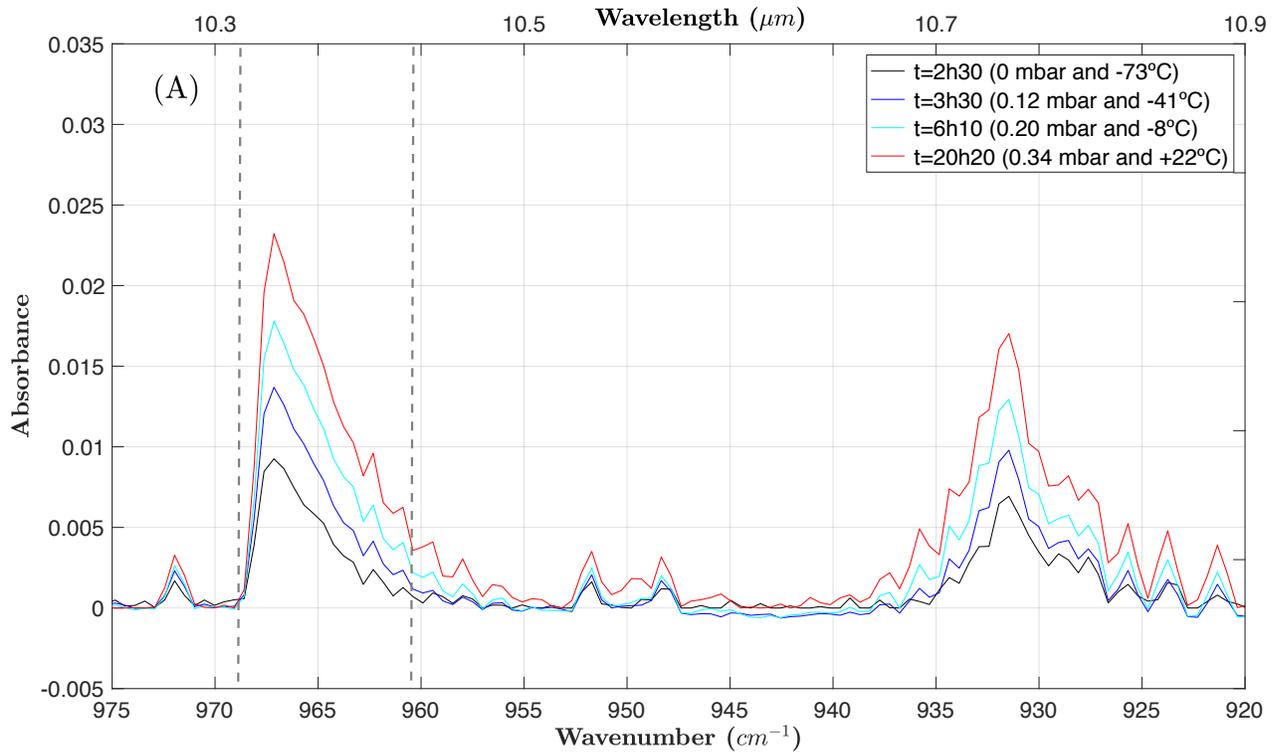

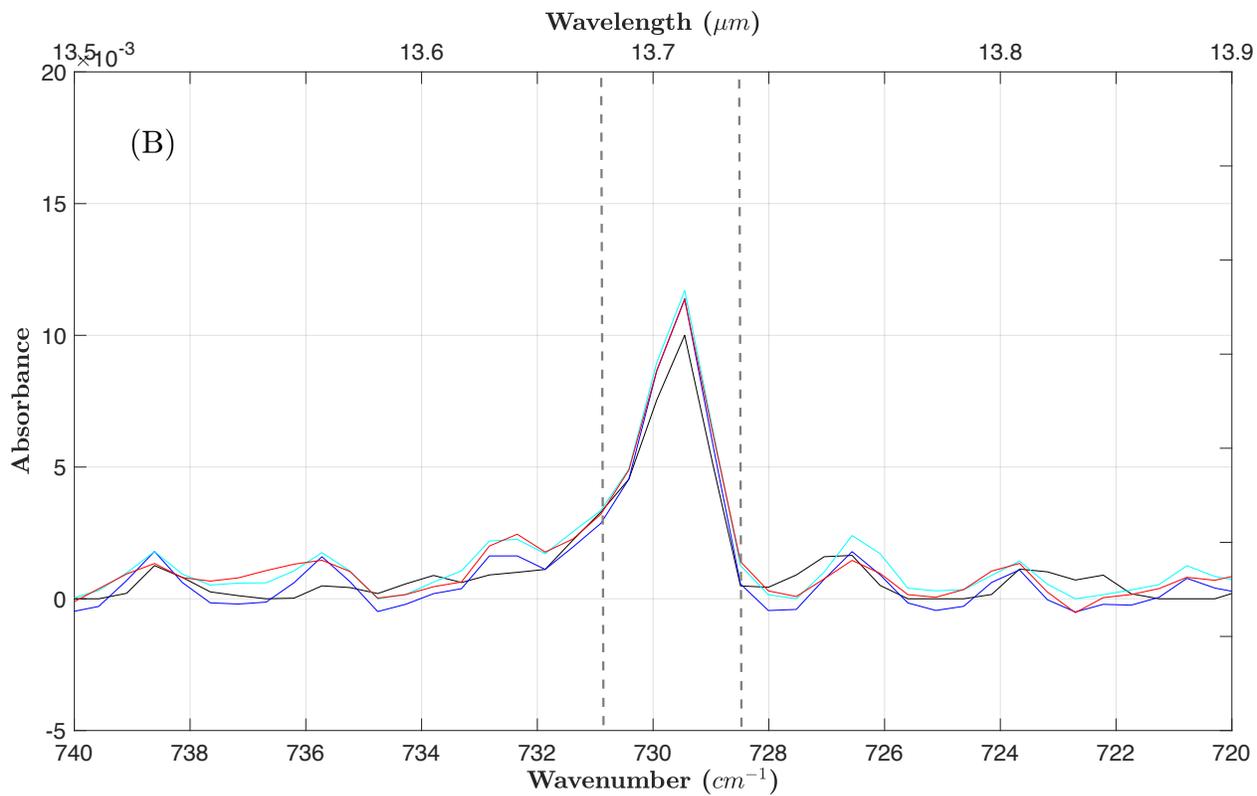
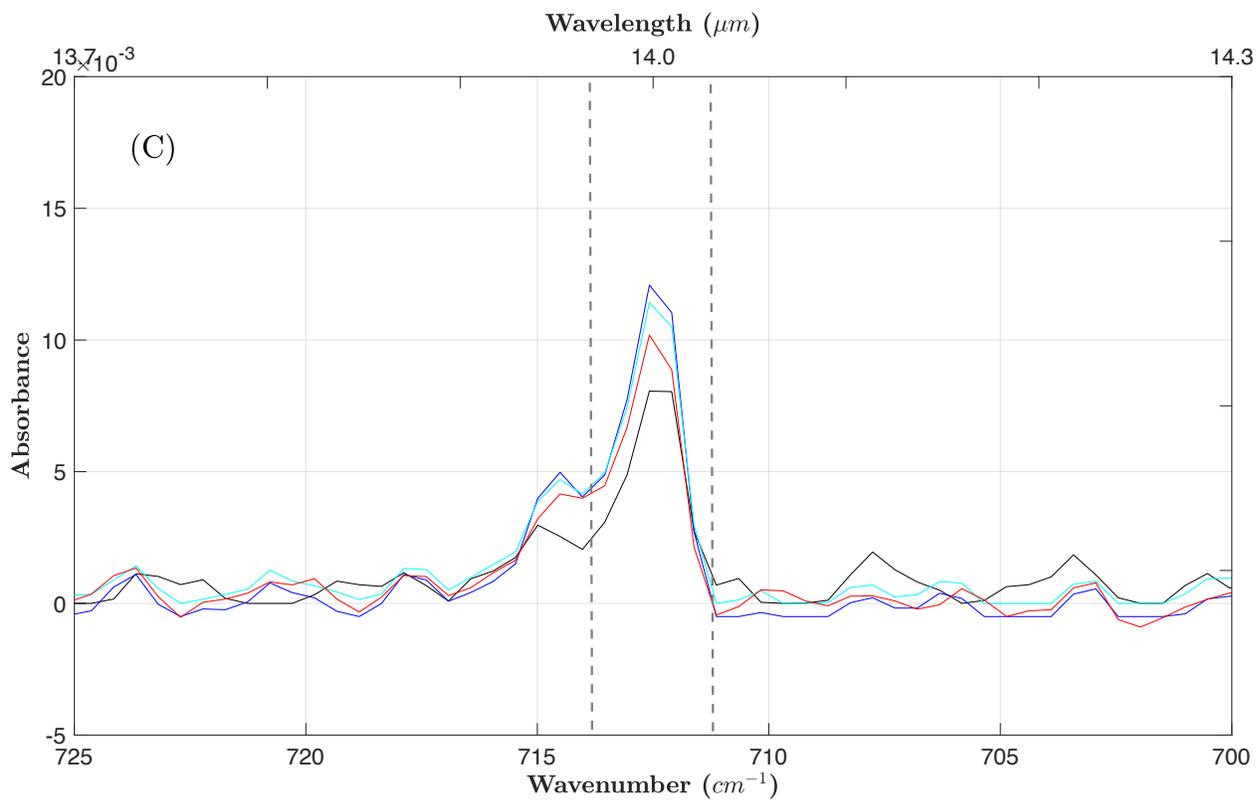

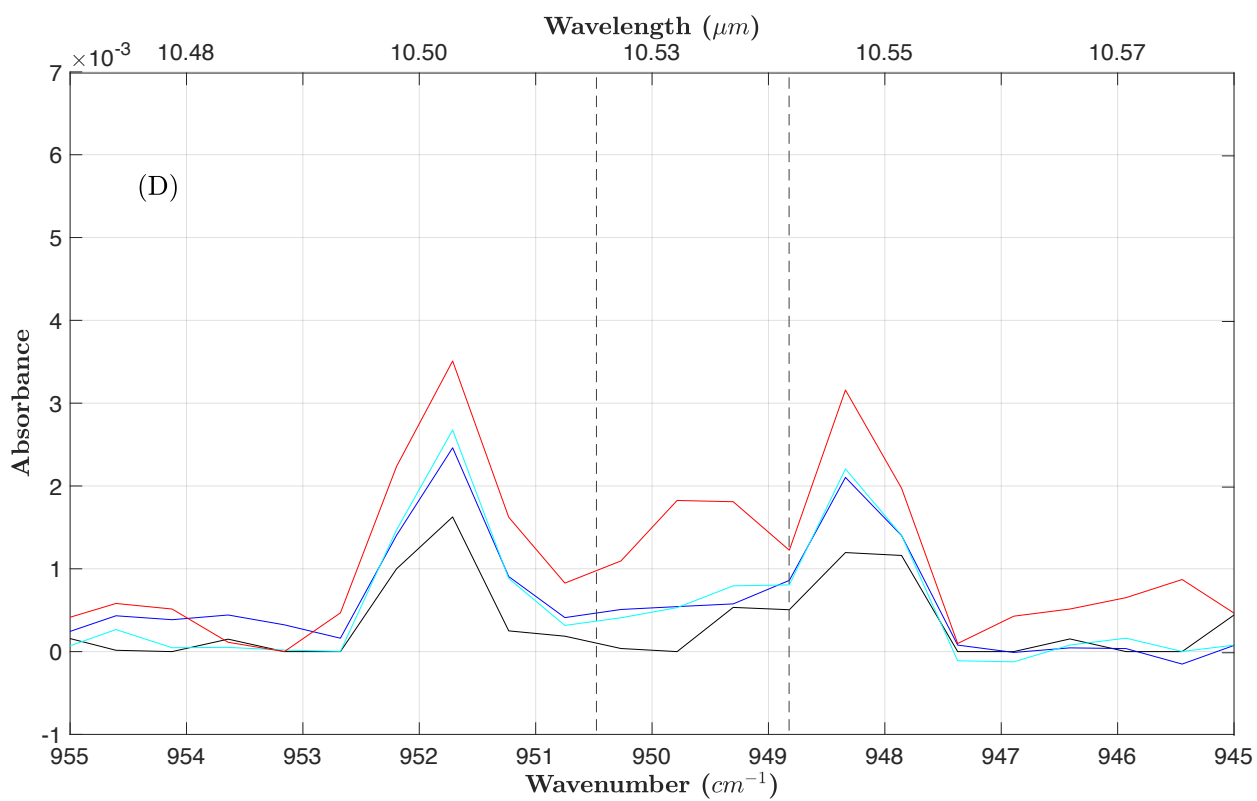

**Fig. 9.** [CH$_4$]$_0$ = 1%. Main absorption bands of (A) NH$_3$ (930 cm$^{-1}$ and 960 cm$^{-1}$ doublet), (B) C$_2$H$_2$ (729.25 cm$^{-1}$), (C) HCN (713 cm$^{-1}$), (D) C$_2$H$_4$ (949.55 cm$^{-1}$). The color code is the same as in Figure 6. The vertical dashed gray lines correspond to the integration band used for the density calculations on either side of the absorption peaks.

**10% CH₄**

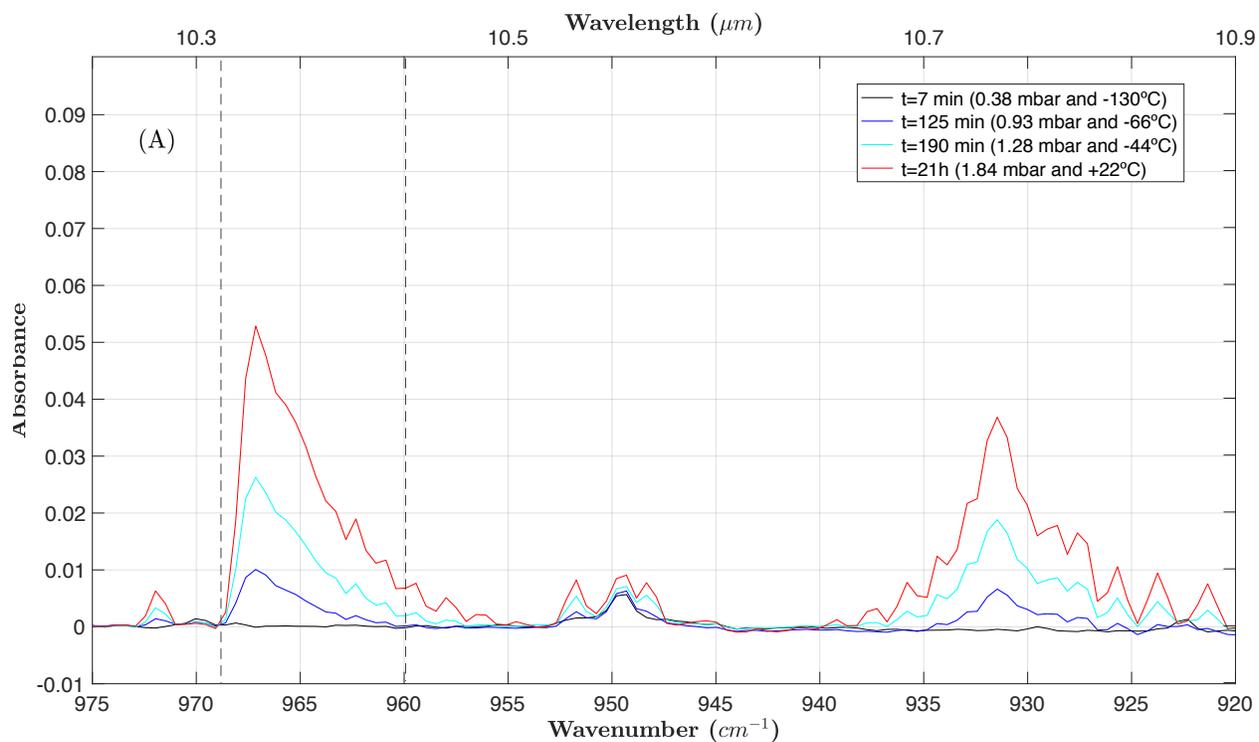

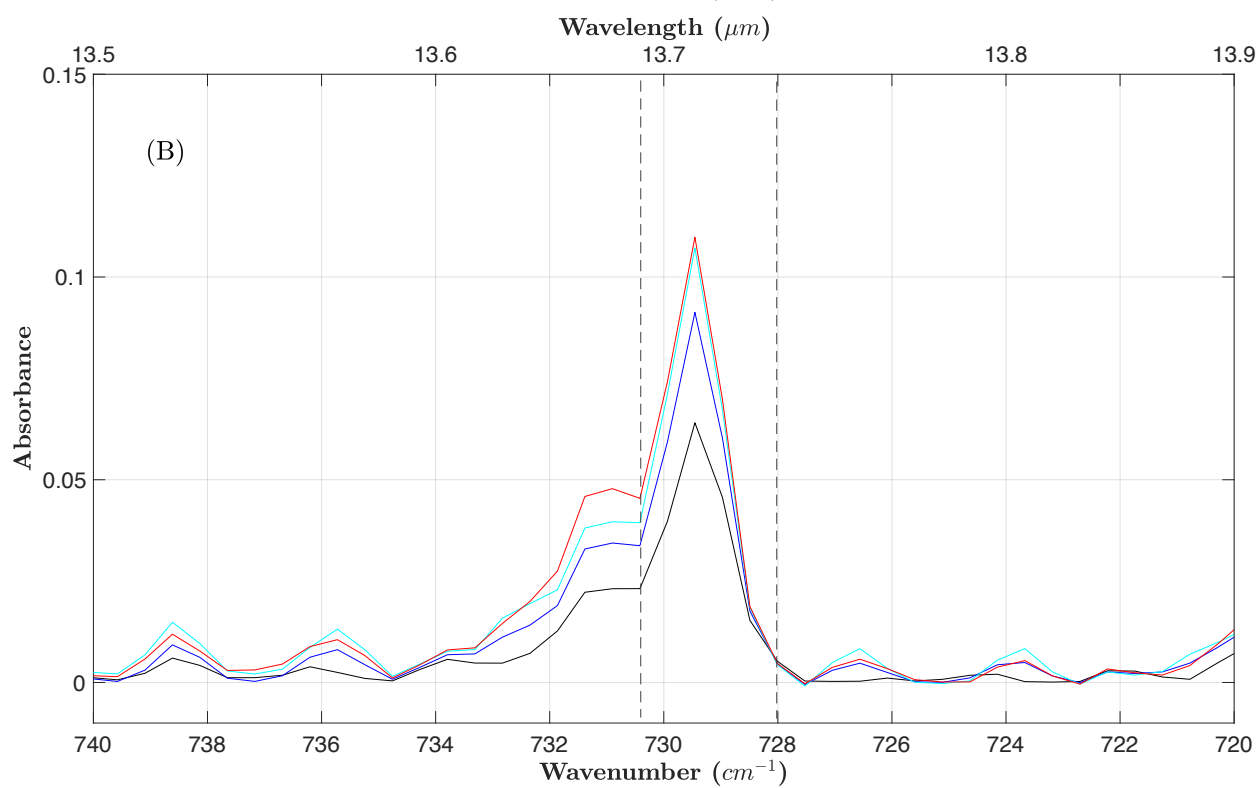

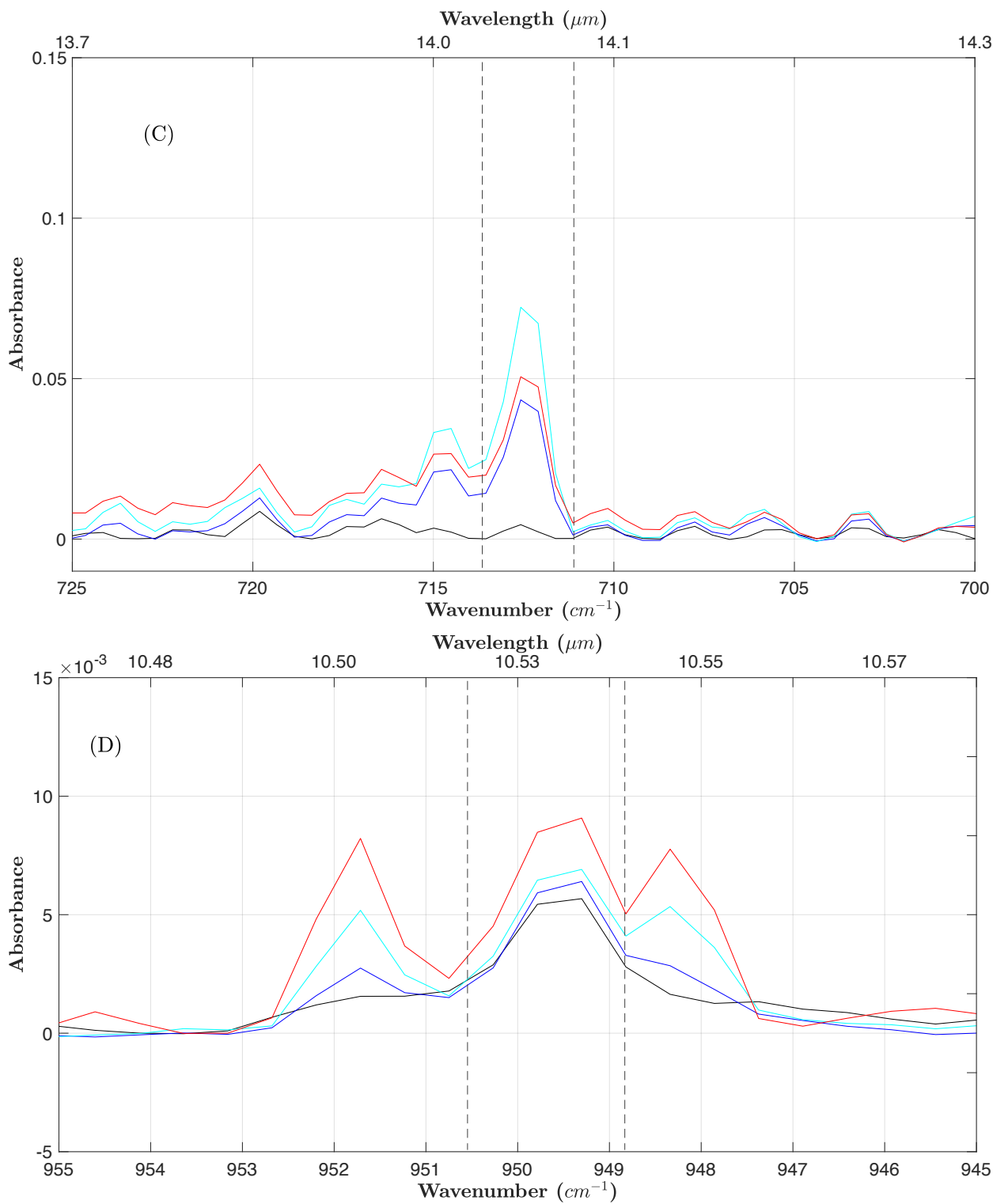

**Fig. 10.** [CH$_4$]$_0$ = 10%. Main absorption bands of (A) NH$_3$ (930 cm$^{-1}$ and 960 cm$^{-1}$ doublet), (B) C$_2$H$_2$ (729.25 cm$^{-1}$), (C) HCN (713 cm$^{-1}$), (D) C$_2$H$_4$ (949.55 cm$^{-1}$). The color code is the same as in Figure 7. The vertical dashed gray lines correspond to the integration band used for the density calculations on either side of the absorption peaks.

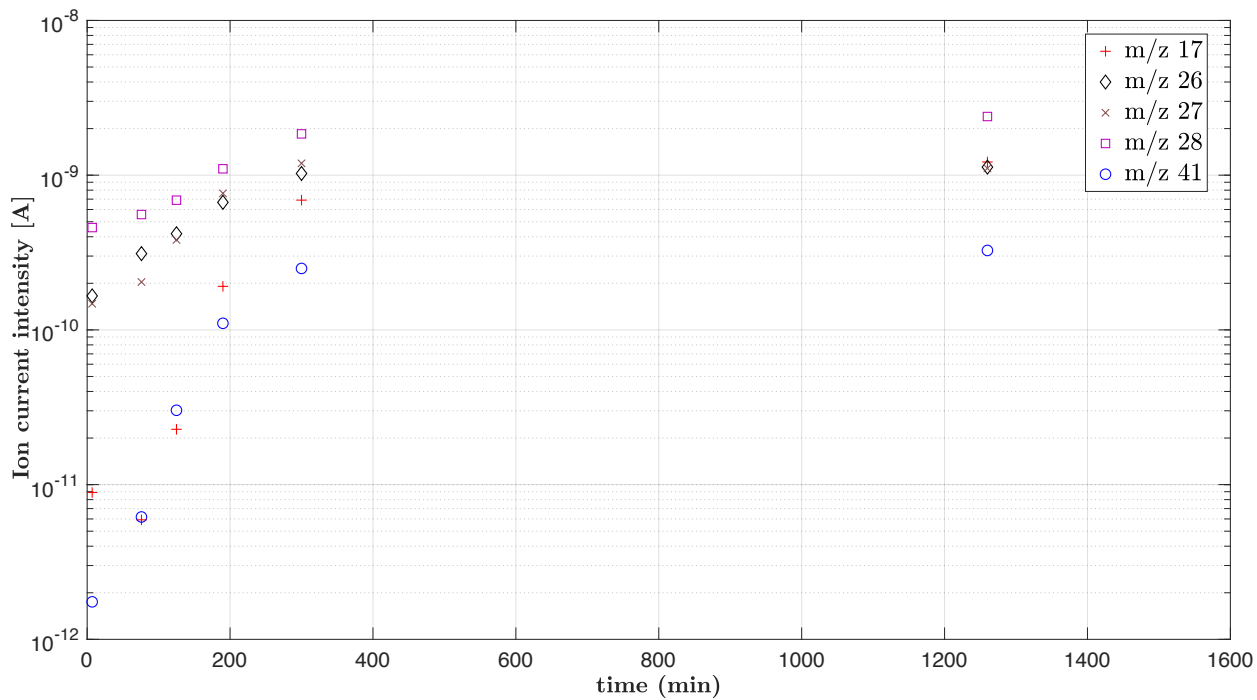

**Fig. 11.** Evolution of the intensities (same time scale as Figure 3) of selected species, $NH_3$ (m/z 17), $C_2H_2$ (m/z 26), HCN (m/z 27), $C_2H_4$ (m/z 28) and $CH_3CN$ (m/z 41) over time.

| Selected species | Fundamental frequency[1] | Main peak (cm$^{-1}$) | FWHM (cm$^{-1}$) | Integrated absorption cross-section $\sigma$ (cm$^2$.molecule$^{-1}$) |
|---|---|---|---|---|
| Ammonia | $\upsilon_2$ | 965 | ± 4.0 | 4.04 x 10$^{-18}$ |
| Acetylene | $\upsilon_5$ | 729.25 | ± 1.25 | 1.92 x 10$^{-16}$ |
| Hydrogen Cyanide | $\upsilon_2$ | 712.3 | ± 1.2 | 4.09 x 10$^{-18}$ |
| Ethylene | $\upsilon_7$ | 949.5 | ± 1.0 | 2.43 x 10$^{-17}$ |

**Table 3.** IR peak locations, fundamental frequencies and integrated wavenumber ranges, absorption cross-sections (cm$^2$.molecule$^{-1}$) for each selected volatile, as analyzed in both methane conditions. For the calculation results (derived from Equ. 6), see the following Tables 4-5. [1]Sharpe et al., 2004 and http://vpl.astro.washington.edu/spectra/allmoleculeslist.htm

| Time/Species | $NH_3$ $(cm^{-3})$ | | | $C_2H_2$ $(cm^{-3})$ | | | HCN $(cm^{-3})$ | | | $C_2H_4$ $(cm^{-3})$ | | |
|---|---|---|---|---|---|---|---|---|---|---|---|---|
| $T_1$ | $5.6 \times 10^{11}$ | $2.8 \times 10^{13}$ | $1.4 \times 10^{14}$ | $2.1 \times 10^{12}$ | $2.4 \times 10^{12}$ | $1.4 \times 10^{12}$ | $4.0 \times 10^{12}$ | $2.5 \times 10^{13}$ | - | - | - | - |
| mean | $5.6 \times 10^{13}$ | | | $2.0 \times 10^{12}$ | | | $9.7 \times 10^{12}$ | | | - | | |
| $T_2$ | $7.0 \times 10^{13}$ | $3.1 \times 10^{14}$ | $4.0 \times 10^{14}$ | $3.1 \times 10^{12}$ | $3.0 \times 10^{12}$ | $2.8 \times 10^{12}$ | $1.3 \times 10^{13}$ | $8.3 \times 10^{13}$ | $1.4 \times 10^{13}$ | - | - | - |
| mean | $2.6 \times 10^{14}$ | | | $3.0 \times 10^{12}$ | | | $3.7 \times 10^{13}$ | | | - | | |
| $T_3$ | $1.6 \times 10^{14}$ | $5.6 \times 10^{14}$ | $6.1 \times 10^{14}$ | $3.1 \times 10^{12}$ | $2.9 \times 10^{12}$ | $3.2 \times 10^{12}$ | $1.7 \times 10^{13}$ | $1.2 \times 10^{14}$ | $2.8 \times 10^{13}$ | - | - | - |
| mean | $4.4 \times 10^{14}$ | | | $3.1 \times 10^{12}$ | | | $5.5 \times 10^{13}$ | | | - | | |
| $T_4$ | $5.2 \times 10^{14}$ | $5.3 \times 10^{14}$ | $8.2 \times 10^{14}$ | $3.7 \times 10^{12}$ | $3.0 \times 10^{12}$ | $3.4 \times 10^{12}$ | $2.1 \times 10^{14}$ | $1.8 \times 10^{14}$ | $1.7 \times 10^{14}$ | - | - | - |
| mean | $6.2 \times 10^{14}$ | | | $3.4 \times 10^{12}$ | | | $1.9 \times 10^{14}$ | | | - | | |
| $T_5$ | $9.6 \times 10^{14}$ | $8.6 \times 10^{14}$ | $1.3 \times 10^{15}$ | $4.4 \times 10^{12}$ | $3.1 \times 10^{12}$ | $4.5 \times 10^{12}$ | $2.0 \times 10^{14}$ | $1.5 \times 10^{14}$ | $1.7 \times 10^{14}$ | - | - | - |
| mean | $1.0 \times 10^{15}$ | | | $4.0 \times 10^{12}$ | | | $1.7 \times 10^{14}$ | | | - | | |
| $T_6$ | $9.6 \times 10^{14}$ | $1.3 \times 10^{15}$ | $1.1 \times 10^{15}$ | $3.4 \times 10^{12}$ | $3.6 \times 10^{12}$ | $2.6 \times 10^{12}$ | $1.7 \times 10^{14}$ | $1.6 \times 10^{14}$ | $1.1 \times 10^{14}$ | - | - | - |
| mean | $1.1 \times 10^{15}$ | | | $3.2 \times 10^{12}$ | | | $1.5 \times 10^{14}$ | | | - | | |

**Table 4.** Calculated molecular densities from IR absorption at $[CH_4]_0 = 1\%$, for three different data sets. $T_{1-6}$ correspond to measurements done at approximately 30 min (-130ºC), 60 min (-85ºC), 120 min (-70ºC), 180 min (-41ºC), 1200 min (+22ºC) and 1440 min (+22ºC), respectively. Each column corresponds to one experiment. Number density calculations are performed with the integrated absorption cross-section of the ExoMol or Hitran databases at the resolution of our experimental spectra (see Table 3). Blank boxes indicate the absence of the species where no number density was derived.

| Time/Species | NH$_3$ (cm$^{-3}$) | | C$_2$H$_2$ (cm$^{-3}$) | | HCN (cm$^{-3}$) | | C$_2$H$_4$ (cm$^{-3}$) | |
|---|---|---|---|---|---|---|---|---|
| T$_1$ | - | - | 2.0 x 10$^{13}$ | 2.2 x 10$^{13}$ | - | - | 8.3 x 10$^{12}$ | 5.5 x 10$^{12}$ |
| mean | - | | 2.1 x 10$^{13}$ | | - | | 6.9 x 10$^{12}$ | |
| T$_2$ | 3.5 x 10$^{14}$ | - | 2.9 x 10$^{13}$ | 3.0 x 10$^{13}$ | 6.5 x 10$^{14}$ | 4.5 x 10$^{15}$ | 1.4 x 10$^{13}$ | 3.3 x 10$^{12}$ |
| mean | 3.5 x 10$^{14}$ | | 3.0 x 10$^{13}$ | | 2.6 x 10$^{15}$ | | 8.7 x 10$^{12}$ | |
| T$_3$ | 8.6 x 10$^{14}$ | 1.6 x 10$^{15}$ | 3.5 x 10$^{13}$ | 4.9 x 10$^{13}$ | 1.1 x 10$^{15}$ | 2.4 x 10$^{15}$ | 1.1 x 10$^{13}$ | 5.9 x 10$^{13}$ |
| mean | 1.2 x 10$^{15}$ | | 4.2 x 10$^{13}$ | | 1.8 x 10$^{15}$ | | 3.5 x 10$^{13}$ | |
| T$_4$ | 2.3 x 10$^{15}$ | 5.0 x 10$^{14}$ | 4.1 x 10$^{13}$ | 3.2 x 10$^{13}$ | 8.5 x 10$^{14}$ | 1.5 x 10$^{15}$ | 2.2 x 10$^{13}$ | 4.3 x 10$^{13}$ |
| mean | 1.4 x 10$^{15}$ | | 3.7 x 10$^{13}$ | | 1.2 x 10$^{15}$ | | 3.3 x 10$^{13}$ | |

**Table 5.** Calculated molecular densities from IR absorption at [CH$_4$]$_0$ = 10%, for two different data sets. T$_{1-4}$ correspond to measurements done at approximately 7 min (-130ºC), 125 min (-66ºC), 190 min (-44ºC) and 21h (+22ºC), respectively. Each column corresponds to one experiment. Number density calculations are performed with the integrated absorption cross-section of the ExoMol or Hitran databases at the resolution of our experimental spectra (see Table 3). Blank boxes indicate the absence of the species where no number density was derived.